\documentclass[a4paper,11pt]{article}

\usepackage{jheppub} 

\usepackage[T1]{fontenc} 

\usepackage{atlasphysics}
\usepackage{multirow}

\usepackage{array}
\newcolumntype{C}{>{\centering\arraybackslash}p{2em}}

\usepackage{preprintcover}  

\PreprintCoverPaperTitle{\boldmath Measurement of $\chicj1$ and $\chicj2$ production with $\rts=7\TeV$ $pp$ collisions at ATLAS}  

\PreprintIdNumber{CERN-PH-EP-2013-202} 

\PreprintCoverAbstract{The prompt and non-prompt production cross-sections for the $\chicj1$ and $\chicj2$ charmonium states are measured in $pp$ collisions at $\rts=7\TeV$ with the ATLAS detector at the LHC using $4.5\,\ifb$ of integrated luminosity. The $\chic$ states are reconstructed through the radiative decay $\chicjgamma$ (with $\Jmumu$) where photons are reconstructed from $\gamma\to\ee$ conversions. The production rate of the $\chicj2$ state relative to the $\chicj1$ state is measured for prompt and non-prompt $\chic$ as a function of $\Jpsi$ transverse momentum. The prompt $\chic$ cross-sections are combined with existing measurements of prompt $\Jpsi$ production to derive the fraction of prompt $\Jpsi$ produced in feed-down from $\chic$ decays. The fractions of $\chicj1$ and $\chicj2$ produced in $b$-hadron decays are also measured.}

\PreprintJournalName{JHEP}  

\clearpage

\title{\boldmath Measurement of $\chicj1$ and $\chicj2$ production with $\rts=7\TeV$ $pp$ collisions at ATLAS}

\author{The ATLAS Collaboration}

\mathchardef\mhyphen="2D

\abstract{The prompt and non-prompt production cross-sections for the $\chicj1$ and $\chicj2$ charmonium states are measured in $pp$ collisions at $\rts=7\TeV$ with the ATLAS detector at the LHC using $4.5\,\ifb$ of integrated luminosity. The $\chic$ states are reconstructed through the radiative decay $\chicjgamma$ (with $\Jmumu$) where photons are reconstructed from $\gamma\to\ee$ conversions. The production rate of the $\chicj2$ state relative to the $\chicj1$ state is measured for prompt and non-prompt $\chic$ as a function of $\Jpsi$ transverse momentum. The prompt $\chic$ cross-sections are combined with existing measurements of prompt $\Jpsi$ production to derive the fraction of prompt $\Jpsi$ produced in feed-down from $\chic$ decays. The fractions of $\chicj1$ and $\chicj2$ produced in $b$-hadron decays are also measured.}

\begin{document} 
\maketitle
\flushbottom


\section{Introduction}

The study of heavy quarkonium production in hadronic collisions offers a unique insight into the dynamics of the strong interaction. Understanding the hadronic production of quarkonium states within quantum chromodynamics (QCD) has been a long-standing challenge, complicated by the presence of several important energy scales~\cite{BBL}. While the production of a heavy quark pair is generally a high-energy process that can be well described perturbatively, the energy scale associated with the evolution of a heavy quark pair into a physical bound state introduces large uncertainties to theoretical predictions.

Understanding the hadronic production of the charmonium states is particularly challenging as the mass of the charm quark is such that the modelling of the bound state as a simple non-relativistic system is less well motivated than for the heavier bottomonium system. Several theoretical approaches have been developed to describe hadronic charmonium production, though the wealth of production and polarisation measurements that now exist are not comprehensively described by any single theoretical approach~\cite{QWG}. However, progress is being made in the calculation of colour-singlet (CS) and colour-octet (CO) production processes at higher perturbative orders and recent calculations provide a good description of the world data on prompt $\Jpsi$ production cross-sections~\cite{BandK}.

The $\chicj J(1P)$ states (with $J=0,1,2$) are the only triplet of $P$-wave states below the open-charm threshold. The spectroscopy of these states is characterised by small hyperfine mass splittings and the branching fractions for the decays $\chicj J \to\Jpsi\,\gamma$ are large for the $J=1,2$ states ($34.4\%$ and $19.5\%$, respectively), while the corresponding branching fraction for the $J=0$ state is significantly lower ($1.3\%$)~\cite{PDG}. The $\chicj J$ states may be produced directly in hadronic collisions or through the decay of higher-mass quarkonium states; these production modes are referred to as prompt. In addition to prompt production, the decay chains of $b$-hadrons can also produce $\chicj J$ states; these production modes are referred to as non-prompt.

The large cross-section for inclusive charmonium production and extensive data samples available at the Large Hadron Collider (LHC) allow the hadronic production of charmonium to be studied in detail. The inclusive production rate of prompt $\Jpsi$ is the most experimentally accessible charmonium production observable at the LHC, with reconstruction of the decay $\Jmumu$ being well suited to the hadronic environment. However, the comparison of experimental measurements with theoretical predictions is complicated by the large feed-down contributions from $\chic$ and $\psi(2S)$ decays. The direct component of the inclusive $\Jpsi$ cross-section, produced in the $pp$ interaction, can be obtained only if these feed-down contributions are precisely quantified. Existing measurements suggest that the contribution to prompt $\Jpsi$ production from $\chic$ decays is around $25\%$~\cite{FeedDown}. An understanding of $\chic$ production is therefore a crucial component of any general description of charmonium production at the LHC. Furthermore, $\chic$ production observables, such as the relative production rates of the $\chicj1$ and $\chicj2$ states, represent sensitive probes of the prompt charmonium production mechanism that can provide information complementary to the study of the $S$-wave states.

The production of charmonium states in $b$-hadron decays can be used as a proxy observable for studying $b$-quark production at the LHC. Theoretical predictions of $b$-quark production can be combined with fragmentation functions and momentum spectra from $H_{b}\to\left(\ccbar\right) X$ decays (where $H_{b}$ denotes a $b$-hadron and $\left(\ccbar\right)$ denotes a charmonium state) extracted from $\ee$ collision data to provide direct predictions for non-prompt charmonium production~\cite{FONLL,FONLL2}. Such predictions have had much success in describing the measurements of non-prompt $\Jpsi$ and $\psi(2S)$ production performed by the LHC experiments~\cite{ATLAS_Jpsi,CMS_Psi2S,LHCb_Jpsi,ALICE1,ALICE2}.

Various aspects of the production of $\chic$ states have been studied at the LHC~\cite{LHCb_Ratio,LHCb_Ratio2,LHCB_ChiJpsiRatio,CMS_Ratio} and at the Tevatron~\cite{CDF_ChiJpsiRatio,CDF_Ratio}; however, measurements of the absolute production cross-sections for prompt $\chic$ and studies of non-prompt $\chic$ production have not been performed previously at the LHC.

This paper presents measurements of the inclusive production of the $\chicj1$ and $\chicj2$ states in $pp$ collisions at a centre-of-mass energy $\rts=7\TeV$ at the LHC. The $\chic$ states are reconstructed through the radiative decays $\chicj J \to \Jpsi\,\gamma$ (with $\Jmumu$), where the photon is reconstructed through its conversion into a positron--electron ($\ee$) pair. Photon conversions reconstructed in the ATLAS inner tracking detector offer the very good mass resolution needed to resolve the $\chicj1$ and $\chicj2$ states individually. The $\chicj0$ production rate is not measured explicitly because the inclusive yield of $\chicj0$ in the data sample used in this analysis is considered to be too small to perform a reliable measurement.

The inclusive production of the $\chicj1$ and $\chicj2$ states is separated experimentally into prompt and non-prompt components and measured differentially in both $\chic$ transverse momentum $\ptchic$ and $\Jpsi$ transverse momentum $\ptjpsi$, within the rapidity region $|\yjpsi|<0.75$. The results obtained as a function of $\ptjpsi$ and $\ptchic$ are presented within the regions $10\leq\ptjpsi<30\GeV$ and $12\leq\ptchic<30\GeV$ respectively. These new measurements are combined with existing measurements of inclusive $\Jpsi$ production \cite{ATLAS_Jpsi} to determine the fraction of prompt $\Jpsi$ produced in feed-down from $\chic$ decays. The ratio $\sigma\left(\chicj2 \right)/\sigma\left(\chicj1 \right)$ is a useful theoretical quantity as it is sensitive to the presence of possible colour-octet contributions to $\chic$ production~\cite{NRQCD}. This cross-section ratio is measured as a function of $\ptjpsi$ for prompt $\chicj1$ and $\chicj2$ production. These measurements are compared to theoretical predictions and to the measurements by other experiments. The branching fraction ${\cal B}(B^{\pm}\to\chicj1 K^{\pm})$ is also measured with the same data sample and event selection.


\section{The ATLAS detector, data and Monte Carlo samples}
\label{Sec:ATLASDataMC}

The ATLAS detector~\cite{ATLAS} is a general-purpose particle physics detector with a cylindrical geometry\footnote{ATLAS uses a right-handed coordinate system with its origin at the nominal interaction point (IP) in the centre of the detector and the $z$-axis along the beam pipe. The $x$-axis points from the IP to the centre of the LHC ring, and the $y$-axis points upward. Cylindrical coordinates $(r,\phi)$ are used in the transverse plane, $\phi$ being the azimuthal angle around the beam pipe. The pseudorapidity is defined in terms of the polar angle $\theta$ as $\eta=-\ln\tan(\theta/2)$.} with forward-backward symmetric coverage in pseudorapidity $\eta$. The detector consists of inner tracking detectors, calorimeters and a muon spectrometer, and has a three-level trigger system. 

The inner tracking detector (ID) is composed of a silicon pixel detector, a semiconductor microstrip detector (SCT) and a transition radiation tracker, which together cover the pseudorapidity range $|\eta| < 2.5$. The ID directly surrounds the beam pipe and is immersed in a $2\,\mathrm{T}$ axial magnetic field generated by a superconducting solenoid. The calorimeter system surrounds the solenoid and consists of a highly granular liquid-argon electromagnetic calorimeter and an iron/scintillator tile hadronic calorimeter. The muon spectrometer (MS) surrounds the calorimeters and consists of three large air-core superconducting magnets (each with eight coils), which generate a toroidal magnetic field. The MS is instrumented in three layers with precision detectors (monitored drift tubes and cathode strip chambers) that provide precision muon tracking covering $|\eta| < 2.7$ and fast trigger detectors (resistive plate chambers and thin gap chambers) covering the range $|\eta| < 2.4$.

The ATLAS trigger is a three-level system consisting of a level-1 trigger implemented in hardware and a software-based two-stage high level trigger (HLT). The data sample used in this analysis is collected with a dimuon trigger. The level-1 muon trigger system identifies regions of interest (RoI) by searching for coincidences between hits in different trigger detector layers within predefined geometrical windows enclosing the paths of muons with a given set of transverse momentum thresholds. The level-1 system also provides a rough measurement of muon position with a spatial granularity of $\Delta\phi \times \Delta\eta \approx 0.1 \times 0.1$. The level-1 RoI then serves as a seed for HLT algorithms that use higher precision MS and ID measurements to reconstruct muon trigger objects. The selection performed by the HLT algorithms is discussed in section~\ref{Sec:EventSelection}.

The measurements presented in this paper are performed with a data sample corresponding to an integrated luminosity of $4.5\,\ifb$ collected during the 2011 LHC proton--proton run at a centre-of-mass energy $\rts = 7\TeV$.  The selected events were collected under stable LHC beam conditions with the detector in a fully operational state.

Four Monte Carlo (MC) simulation samples are used to estimate the photon conversion reconstruction efficiency and to characterise the modelling of the $\chic$ signal components used in the fitting procedure. The samples consist of simulated $\chicj1\to\Jpsi\,\gamma \to \mumu\gamma$ and $\chicj2\to\Jpsi\,\gamma \to \mumu\gamma$ decays produced either directly in $pp$ interactions or through the process $pp\to\bbbar X\to\chic \,X^{\prime}$. All samples are generated with {\sc Pythia 6}~\cite{Pythia6} and use the ATLAS 2011 MC underlying event and hadronisation model tuning~\cite{MC11}. The response of the ATLAS detector is simulated using {\sc Geant4}~\cite{GEANT4,ATLASMC}. The events are reconstructed using the same algorithms used to process the data. Each $\chic$ event is overlaid with a number of minimum-bias $pp$ events such that the overall distribution of additional $pp$ interactions due to pile-up in data events is accurately described by the simulated samples.


\section{\boldmath Event and $\chic$ candidate selection}
\label{Sec:EventSelection}

The selected events passed a dimuon trigger in which the HLT identified two oppositely charged muons, each with transverse momentum $\pt>4\GeV$. The HLT fits the two candidate muon tracks to a common vertex and a very loose requirement on the vertex $\chi^{2}$ is imposed. Finally, a broad dimuon invariant mass cut ($2.5 < m\left(\mumu\right) < 4.0\GeV$) is applied to select dimuon candidates consistent with $\Jmumu$ decays. Each event passing the trigger selection is required to contain at least one reconstructed $pp$ collision vertex formed from at least three reconstructed tracks with $\pT> 400\MeV$. 

In the offline analysis, muon candidates are formed from reconstructed ID tracks matched to tracks reconstructed in the MS. Each muon track is required to be reconstructed from at least six SCT hits and at least one pixel detector hit. For the low-$\pt$ muons (typically below $20\GeV$) produced in $\Jmumu$ decays, the track parameters measured in the ID provide more accurate measurements than the MS as they are not affected by muon energy loss in the calorimeters. Therefore, the ID measurements alone are used to reconstruct the momentum of muon candidates. Events are required to contain at least one pair of oppositely charged muons, each with transverse momentum $\ptmu>4\GeV$ and $|\etamu|< 2.3$ (the region where the trigger and reconstruction efficiencies are optimal). Each muon pair is fitted to a common vertex and a very loose vertex quality requirement (fully efficient for genuine $\Jmumu$ decays) is imposed. The dimuon pair is considered a $\Jmumu$ candidate if the invariant mass of the pair, calculated from the track parameters recalculated by the vertex fit, satisfies $2.95 < m\left(\mumu\right) < 3.25\GeV$. The event and $\Jmumu$ candidate are retained for further analysis if the two muon candidates reconstructed offline are consistent with the objects reconstructed by the HLT (matched within $\Delta R = \sqrt{(\Delta\phi)^{2} + (\Delta\eta)^{2}} < 0.01$). The rapidity of the $\Jpsi$ candidate reconstructed offline is required to satisfy $|\yjpsi|<0.75$. This selection retains only the candidates reconstructed within the region of the detector with the optimal dimuon mass resolution, which is necessary to reliably resolve the individual $\chicj J$ states.

Photon conversions are reconstructed from pairs of oppositely charged ID tracks whose helices touch when projected onto the transverse plane. Tracks must be reconstructed with transverse momentum $\pT > 400\MeV$ and $|\eta|<2.3$ and contain at least six SCT hits. Track pairs consistent with the photon conversion hypothesis are fitted to a common vertex with their opening angle constrained to be zero. The vertex fit is required to converge with $\chi^{2}$ per degree of freedom less than five. To reject fake conversions from $\pizero\to\ee\gamma$ decays and other promptly produced track pairs, the radial displacement $r$ of the reconstructed conversion vertex with respect to the $z$-axis is required to satisfy $40 < r < 150\,\mathrm{mm}$. This fiducial region includes the three layers of silicon pixels in the ID, and so selects conversions occurring within these silicon pixel layers and their associated service structures. The efficiency for reconstructing converted photons from only ID tracks decreases significantly beyond the upper limit of $150\,\mathrm{mm}$. In this analysis, no information from the calorimeters is used in the reconstruction of photon conversions. The reconstructed momentum of the converted photon is calculated from $\ee$ track parameters recalculated by the vertex fit with the invariant mass constrained to be zero. Reconstructed converted photons are required to have transverse momentum $\ptgamma > 1.5\GeV$ and $|\etagamma|<2.0$.

Candidate $\chic\to\Jpsi\,\gamma\to\mumu\gamma$ decays are selected by associating a reconstructed photon conversion with a candidate $\Jmumu$ decay. To reject combinations of $\Jmumu$ decays and photons not consistent with a $\,\chic$ decay, the impact parameter of the converted photon with respect to the dimuon vertex is required to be less than $5\,\mathrm{mm}$. This requirement has a negligible inefficiency for genuine $\chic$ decays.


\section{Cross-section determination}
\label{Sec:CrossSection}
The distribution of mass difference $\Delta m = m\left(\mumu\gamma\right) - m\left(\mumu\right)$ is used to distinguish the $\chicj1$ and $\chicj2$ states. This distribution is used in place of the three-body invariant mass as some partial cancellation of contributions from the dimuon mass resolution is achieved, resulting in improved overall mass resolution. Non-prompt $\chic$ candidates produced in the decays of $b$-hadrons can be distinguished experimentally from prompt $\chic$ candidates (produced in the primary $pp$ interaction) with the pseudo-proper decay time distribution $\tau$. The pseudo-proper decay time $\tau$ is defined as

\begin{equation}
\tau = \frac{L_{xy}\cdot m_{\Jpsi}}{p_{\mathrm{T}}}\,,\nonumber
\label{Eqn:Tau}
\end{equation}

\noindent where $m_{\Jpsi}$ is the world-average $\Jpsi$ mass, $\pt$ is the transverse momentum of the $\Jpsi$ candidate and $L_{xy}$ is the distance between the primary $pp$ interaction vertex and the $\Jmumu$ decay vertex in the transverse plane. The primary $pp$ interaction vertex, defined as the vertex with the highest track $\sum p^{2}_{\mathrm{T}}$, is used to calculate $L_{xy}$ on a per-candidate basis.

The differential $\chicj1$ and $\chicj2$ cross-sections for prompt and non-prompt production in a given bin of $\ptchic$ are measured from,

\begin{equation}
\frac{\mathrm{d}\sigma\left(\chicj J\right)}{\mathrm{d}\ptchic}\cdot{\cal B}\left(\chicj J \to\Jpsi\,\gamma\right)\cdot{\cal B}\left(\Jmumu\right) = \frac{N_{J}}{{\cal L}\cdot\Delta\ptchic}\,,\nonumber
\label{Eqn:CrossSectionChi}
\end{equation}

\noindent where ${\cal L}$ is the integrated luminosity, $\Delta\ptchic$ is the bin width in $\ptchic$ and $N_{J}$ is the acceptance- and efficiency-corrected fitted $\chicj J$ signal yield for a given bin in $\ptchic$. The same formula is also used to deduce the differential cross-sections measured as a function of  $\Jpsi$ transverse momentum with $\ptchic$ replaced by $\ptjpsi$.

To obtain corrected prompt and non-prompt $\chicj1$ and $\chicj2$ yields, a weight is first determined for each $\chic$ candidate to correct for experimental efficiency and detector acceptance. Corrected yields are then extracted from a simultaneous two-dimensional unbinned maximum likelihood fit to the weighted mass difference $\Delta m$ and pseudo-proper decay time $\tau$ distributions in bins of $\ptchic$ and $\ptjpsi$. The differing production kinematics for prompt and non-prompt $\chicj1$ and $\chicj2$ result in a different detector acceptance for each yield. To account for these differences, the fit procedure is performed once for each measured yield, with the data weighted differently each time to accurately correct that particular yield. The correction weight $w$ for each $\chic$ candidate is calculated as

\begin{equation}
w^{-1} = {\cal A}\cdot\epsilon_{\mathrm{trig}}\cdot\epsilon_{\mathrm{dimuon}}\cdot\epsilon_{\gamma}\,,\nonumber
\label{Eqn:Weight}
\end{equation}

\noindent where ${\cal A}$ is the detector acceptance, $\epsilon_{\mathrm{trig}}$ is the trigger efficiency, $\epsilon_{\mathrm{dimuon}}$ is the dimuon reconstruction efficiency and $\epsilon_{\gamma}$ is the photon conversion reconstruction efficiency.


\subsection{Acceptance}
\label{Sec:Acc}

The detector acceptance ${\cal A}$ is defined as the probability that final-state decay products in a $\chic\to\Jpsi\,\gamma\to\mumu\gamma$ decay fall within the fiducial region defined by $\ptmu>4\GeV$ and $|\etamu| < 2.3$ for muons and $\ptgamma > 1.5\GeV$ and $|\etagamma|<2.0$ for photons. The acceptance depends strongly on the angular distributions of the decay products in their respective decay frames. The form of these angular distributions is a function of the spin-alignment of the $\chicj J$ with respect to a given axis.

The angular distribution of the $\mu^{+}$ in the $\Jpsi$ rest frame for inclusive production is described by

\begin{equation}
\frac{\mathrm{d}^{2}N}{\mathrm{d}\cos{\theta}d\phi} \propto \frac{1}{3+\lambda_{\theta}}\left[ 1 + \lambda_{\theta}\cos^2{\theta} + \lambda_{\phi}\sin^{2}{\theta}\cos{2\phi} + \lambda_{\theta\phi}\sin{2\theta}\cos{\phi} \right]\,,
\label{Eqn:AngMuPlus}
\end{equation}

\noindent where $\theta$ and $\phi$ are the polar and azimuthal angles, respectively, between the $\mu^{+}$ and the chosen spin-alignment axis~\cite{Faccioli_ChiPol}. Similarly, the angular distribution of the $\Jpsi$ in the $\chicj J$ rest frame is described by

\begin{equation}
\frac{\mathrm{d}^{2}N}{\mathrm{d}\cos{\Theta}d\Phi} \propto \frac{1}{3+\lambda_{\Theta}}\left[ 1 + \lambda_{\Theta}\cos^2{\Theta} + \lambda_{\Phi}\sin^{2}{\Theta}\cos{2\Phi} + \lambda_{\Theta\Phi}\sin{2\Theta}\cos{\Phi} \right]\,,
\label{Eqn:AngJpsi}
\end{equation}

\noindent where $\Theta$ and $\Phi$ are the polar and azimuthal angles, respectively, between the $\Jpsi$ and the chosen spin-alignment axis. This distribution is valid for $\chicj1$, and is also valid for $\chicj2$ if contributions from higher-order multipoles to the radiative transition $\chicj 2 \to \Jpsi\,\gamma$ are neglected (the $\mathrm{E1}$ approximation). This approximation is motivated by the fact that the current experimental data suggest that higher-order multipoles in both $J=1$ and $J=2$ radiative decays represent less than $10\%$ of the total amplitude~\cite{PDG}. The values of the $\lambda$ parameters in equations\,(\ref{Eqn:AngMuPlus}) and\,(\ref{Eqn:AngJpsi}) are determined by the $\Jpsi$ and $\chic$ spin-alignments, respectively.

For the radiative decay $\chic\to\Jpsi\,\gamma$, it has been shown that the angular distribution of the $\mu^{+}$ in the $\Jpsi$ rest frame and the $\Jpsi$ in the $\chic$ rest frame are identical (i.e. $\lambda_{\Theta}=\lambda_{\theta}$, $\lambda_{\Phi}=\lambda_{\phi}$, $\lambda_{\Theta\Phi}=\lambda_{\theta\phi}$) if one measures the angles $\Theta,\Phi$ and $\theta,\phi$ with respect to parallel spin-alignment axes~\cite{Faccioli_ChiPol}. Furthermore, this choice of axes reduces the dependence of the $\lambda$ parameters in equation\,(\ref{Eqn:AngMuPlus}) on the higher-order multipole contributions to the radiative transitions~\cite{Faccioli_ChiPol}. In this analysis, the helicity frame is used and the spin-alignment axis is defined as the $\chicj J$ line of flight in the laboratory frame.

Scenarios are identified (four for $\chicj1$ and five for $\chicj2$) that span the allowed $\lambda$ parameter space and give rise to the most extreme variations in the acceptance. These are detailed in table~\ref{Table:Pol}. The scenarios correspond to the pure helicity eigenstates of the $\chicj1$ and $\chicj2$ states along with two scenarios with an azimuthal anisotropy ($\mathrm{AZ+}$ and $\mathrm{AZ-}$). These scenarios are used to calculate the uncertainty envelope associated with spin-alignment effects. The central value for the acceptance is calculated by assuming that the $\chicj J$ are unpolarised, with isotropic angular distributions for the $\chic$ and $\Jpsi$ decays ($\lambda_{\theta}=\lambda_{\phi}=\lambda_{\theta\phi}=0$). Spin-alignment scenarios with non-zero values for $\lambda_{\theta\phi}$ are found to result in relative changes in the acceptance below any of the other scenarios considered.

\begin{table}[h]
\begin{center}
\begin{tabular}{| c | c | c | c | c |}
\hline
& Label & $\lambda_{\theta}$ & $\lambda_{\phi}$ & $\lambda_{\theta\phi}$ \\
\hline                        
\multirow{4}{*}{$\chicj1$} & Isotropic & $0$ & $0$ & $0$ \\
 & Helicity $0$ & $+1$ & $0$ & $0$ \\
 & Helicity $\pm1$ & $-1/3$ & $0$ & $0$ \\
 & $\mathrm{AZ+}$ & $-1/3$ & $+1/3$ & $0$ \\
 & $\mathrm{AZ-}$ & $-1/3$ & $-1/3$ & $0$ \\
\hline
\multirow{5}{*}{$\chicj2$} & Isotropic & $0$ & $0$ & $0$ \\
 & Helicity $0$ & $-3/5$ & $0$ & $0$ \\
 & Helicity $\pm1$ & $-1/3$ & $0$ & $0$ \\
 & Helicity $\pm2$ & $+1$ & $0$ & $0$ \\
 & $\mathrm{AZ+}$ & $+1/5$ & $+1/\sqrt{5}$ & $0$ \\
 & $\mathrm{AZ-}$ & $+1/5$ & $-1/\sqrt{5}$ & $0$ \\
\hline   
\end{tabular}
\end{center}

\caption{The set of $\chicj1$ and $\chicj2$ spin-alignment scenarios studied.}
\label{Table:Pol}
\end{table}

Two-dimensional acceptance maps binned in $\ptchic$ and $|\ychic|$ are derived for each spin-alignment scenario using a large sample of generator-level MC events. The angular distributions of the decays $\chic\to\Jpsi\,\gamma$ and $\Jmumu$ are generated according to Equations~\ref{Eqn:AngMuPlus} and \ref{Eqn:AngJpsi} with the $\lambda$ parameters shown in table~\ref{Table:Pol}. Figure~\ref{Fig:Acc} shows the acceptance map for $\chicj1\to\Jpsi\,\gamma\to\mumu\gamma$ decays calculated with isotropic decay angular distributions (consistent with unpolarised $\chic$ production). The acceptance is significantly reduced at high rapidity due to the fiducial cut $|\etagamma|<2.0$. Since $\Jmumu$ candidates are only reconstructed within $|\yjpsi| < 0.75$, none of the converted photons associated with the reconstructed $\chic$ candidates approaches $|\etagamma|\approx2.0$ and the analysis is not sensitive to reduced acceptance in this region. The acceptance also decreases significantly towards low $\ptchic$ and is negligible for $\ptchic < 10\GeV$.

\begin{figure}
\begin{center}
\includegraphics[width=0.7\textwidth]{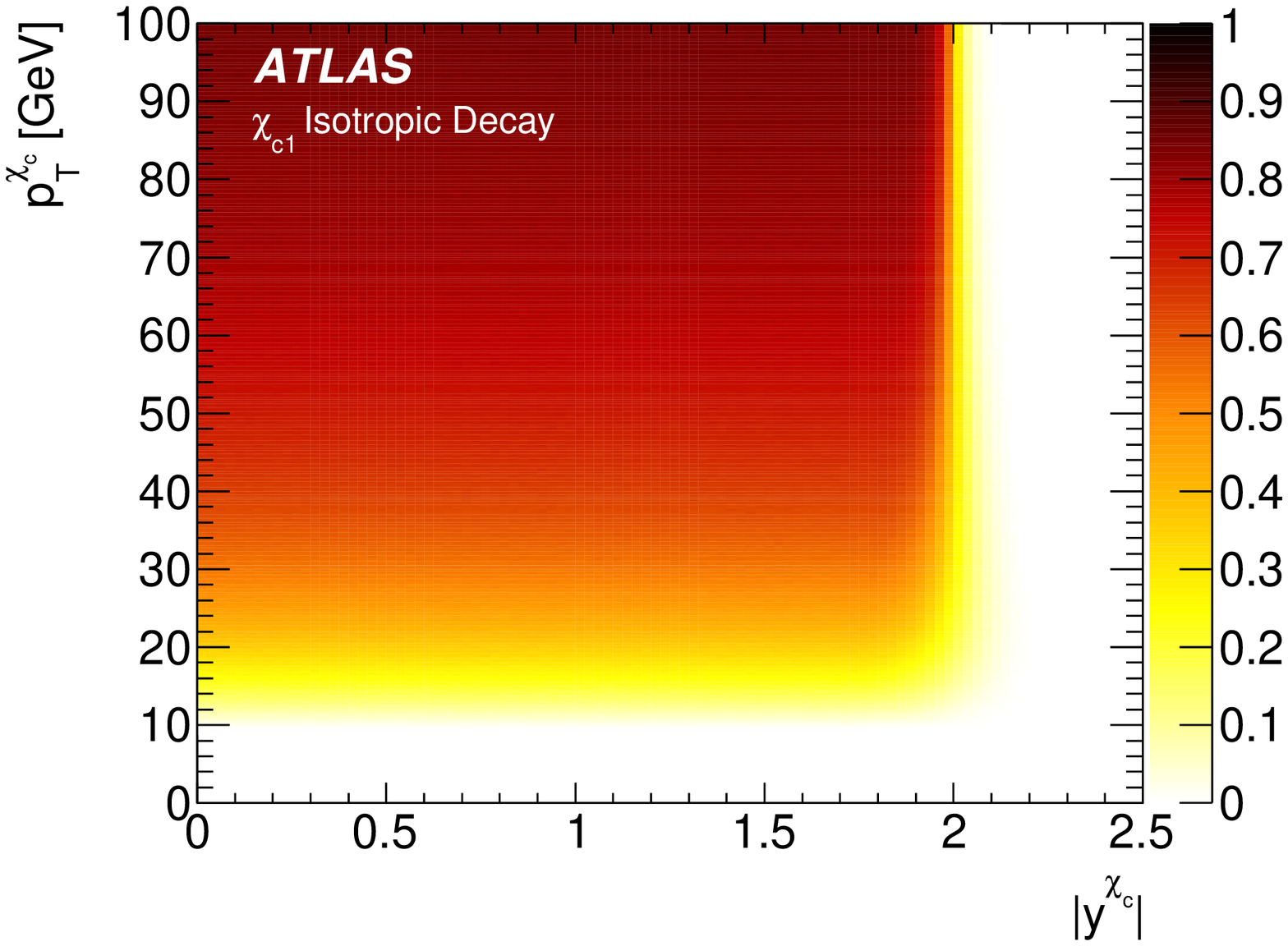}
\end{center}
\caption{The acceptance for $\chicj1\to\Jpsi\,\gamma\to\mumu\gamma$ decays as a function of $\ptchic$ and $|\ychic|$.}
\label{Fig:Acc}
\end{figure}

The cross-section measurements binned in $\ptchic$ are corrected directly with acceptance maps as described above. The acceptance correction for the measurements binned in $\ptjpsi$ is sensitive to the $\ptchic$ spectrum used as input to the simulation and must be calculated with an alternative approach. Acceptance corrections are derived for each measured $\ptjpsi$ bin with the same simulation used to generate the acceptance maps, but with the $\chic$ decays generated with a transverse momentum spectrum taken from a fitted parameterisation of the differential cross-sections measured as a function of $\ptchic$ and presented in this paper.


\subsection{Trigger efficiency}
\label{Sec:Trig}

The dimuon trigger efficiency is determined from $\Jmumu$ and $\Upsilon\to\mumu$ decays in data using the method described in ref.~\cite{ATLAS_Upsilon}. The dimuon trigger efficiency $\epsilon_{\mathrm{trig}}$ is defined as the efficiency with which the trigger system can select events that pass the full offline dimuon selection. The method factorises the total efficiency into three parts,

\begin{equation}
\epsilon_{\mathrm{trig}} = \epsilon_{\mathrm{RoI}}\left(p^{\mu}_{\mathrm{T}1},q_{1}\cdot\eta^{\mu}_{1}\right)\cdot\epsilon_{\mathrm{RoI}}\left(p^{\mu}_{\mathrm{T}2},q_{2}\cdot\eta^{\mu}_{2}\right)\cdot c_{\mumu}\left(\Delta R, |y(\mumu)|\right)\,,\nonumber
\label{Eqn:TrigEff}
\end{equation}

\noindent where $\epsilon_{\mathrm{RoI}}$ is the efficiency with which the trigger system can identify a single muon with transverse momentum $\ptmu$ and charge-signed pseudorapidity $q\cdot\eta^{\mu}$ as an RoI. Charge-signed pseudorapidity is used to account for the charge asymmetry in single-muon triggering due to the toroidal magnetic field. The factor $c_{\mumu}$ is present to correct for inefficiencies associated with the dimuon selection of the trigger. These include vertexing and dimuon charge requirements and effects associated with the finite size of the muon RoI. The factor $c_{\mumu}$ and the single-muon efficiency $\epsilon_{\mathrm{RoI}}$ are determined using a sample of $\Jmumu$ decays selected with a single-muon trigger with an $18\GeV$ transverse momentum threshold. The efficiency $\epsilon_{\mathrm{RoI}}$ is derived in two-dimensional bins of $p^{\mu}_{\mathrm{T}}$ and $q\cdot\eta^{\mu}$ from the fraction of fitted $\Jmumu$ decays that pass the high-$\pt$ single-muon trigger and that also pass the dimuon trigger selection (corrected for dimuon effects with $c_{\mumu}$). The average dimuon trigger efficiency for $\Jmumu$ decays with $10\leq\ptjpsi<30\GeV$ is between $50\%$ to $60\%$ for the fiducial region studied.


\subsection{Muon reconstruction efficiency}

The efficiency to reconstruct and identify both muons from the decay $\Jmumu$ is described by the equation

\begin{equation}
\epsilon_{\mathrm{dimuon}} = \epsilon_{\mathrm{trk}}\left(p^{\mu}_{\mathrm{T}1},\eta^{\mu}_{1}\right)\cdot\epsilon_{\mathrm{trk}}\left(p^{\mu}_{\mathrm{T}2},\eta^{\mu}_{2}\right)\cdot\epsilon_{\mu}\left(p^{\mu}_{\mathrm{T}1},q_{1}\cdot\eta^{\mu}_{1}\right)\cdot\epsilon_{\mu}\left(p^{\mu}_{\mathrm{T}2},q_{2}\cdot\eta^{\mu}_{2}\right)\,.\nonumber
\label{Eqn:MuonEff}
\end{equation}

\noindent The quantity $\epsilon_{\mathrm{trk}}$ is the reconstruction efficiency for muon tracks subject to the ID track selection described in section~\ref{Sec:EventSelection} and is determined to be $(99\pm1)\%$ over the full kinematic region studied~\cite{ATLAS_Upsilon}. The single-muon identification efficiency $\epsilon_{\mu}$ is derived from an analysis of a sample of $\Jmumu$ decays in data, unbiased by any muon selection criteria as described in ref.~\cite{ATLAS_Upsilon}. The muon identification efficiency reaches a plateau of around $98\%$ for muons with $\ptmu>8\GeV$.

The efficiency of the dimuon invariant mass requirement $2.95 < m\left(\mumu\right) < 3.25\GeV$ discussed in section~\ref{Sec:EventSelection} is estimated to be $(99.0\pm0.5)\%$ by performing fits to the $m\left(\mumu\right)$ distribution of an inclusive sample of $\Jmumu$ decays reconstructed within the same fiducial region used in the analysis.


\subsection{Photon conversion reconstruction efficiency}
\label{Sec:ConvEff}
The total photon conversion reconstruction efficiency $\epsilon_{\gamma}$ is determined from the MC-simulated $\chic\to\Jpsi\,\gamma$ event samples described in section~\ref{Sec:ATLASDataMC}. The decay products in the simulated samples of radiative $\chic$ decays are propagated through the ATLAS detector simulation. The description of the ID material distribution in the MC simulation samples is checked by comparing the distributions of various conversion observables (including conversion vertex fit $\chi^{2}$, vertex position and kinematic variables) measured in data and MC simulation. The distributions are found to agree well and no significant discrepancies are observed. Studies of the ID material description in the MC simulation with secondary hadronic interactions also agree well with data~\cite{Material}.

The total photon conversion reconstruction efficiency $\epsilon_{\gamma}$ is factorised into two parts,

\begin{equation}
\epsilon_{\gamma} = P_{\mathrm{conv}}\left(\etagamma\right)\cdot\epsilon_{\mathrm{conv}}\left(\ptgamma,|\etagamma|\right)\,,\nonumber
\label{Eqn:ConvEff}
\end{equation}

\noindent where $P_{\mathrm{conv}}$ is the probability that a photon converts within the fiducial region for conversions ($40 < r < 150\,\mathrm{mm}$) and $\epsilon_{\mathrm{conv}}$ is the converted-photon reconstruction efficiency. The conversion probability $P_{\mathrm{conv}}$ is derived from the ratio of the number of generated photons from radiative $\chic$ decays that convert in the fiducial region of the ID to the total number of photons from radiative $\chic$ decays. The calculation of $P_{\mathrm{conv}}$ is performed in bins of $\etagamma$ to account for changes in the detector material traversed for photons of different pseudorapidity. No significant dependence of $P_{\mathrm{conv}}$ on photon energy is observed for photons within the kinematic fiducial region ($\ptgamma > 1.5\GeV$). The conversion probability varies from around $7\%$ to $11\%$ within the fiducial region studied.

The reconstruction efficiency for converted photons, $\epsilon_{\mathrm{conv}}$, is determined from the equation

\begin{equation}
\epsilon_{\mathrm{conv}} = N^{\gamma}_{\mathrm{reco}}/N^{\gamma}_{\mathrm{conv}}\,,\nonumber
\label{Eqn:ConvRecoEff}
\end{equation}

\noindent where $N^{\gamma}_{\mathrm{conv}}$ is the number of simulated photons from radiative $\chic$ decays that convert in the fiducial region of the ID and $N^{\gamma}_{\mathrm{reco}}$ is the number of reconstructed photon conversions. This ratio is calculated in bins of $\ptgamma$ and $|\etagamma|$ (no significant asymmetry in photon pseudorapidity is observed) with a method that is verified to account correctly for experimental resolution in $\ptgamma$. The conversion reconstruction efficiency is around $15\%$ at $\ptgamma = 1.5\GeV$ and approaches a plateau of approximately $45\%$ for $\ptgamma > 5.0\GeV$. 


\subsection{Extraction of corrected yields}

Corrected $\chicj J$ yields are extracted by performing a simultaneous fit to the mass difference \\$\Delta m = m\left(\mumu\gamma\right)\nobreak - m\left(\mumu\right)$ and pseudo-proper decay time $\tau$ distributions of weighted $\chic$ candidates. Separate unbinned maximum likelihood fits are performed in bins of both $\Jpsi$ candidate transverse momentum $\ptjpsi$ and $\chic$ candidate transverse momentum $\ptchic$. The fits are performed within the $\chic$ signal region of $0.2 < \Delta m < 0.7\GeV$ and no restriction on $\tau$ is applied. The full probability density function (pdf) used to perform the fits has the form,

\begin{equation}
F\left(\Delta m,\tau,\delta\tau\right) = f_{\mathrm{sig}}\cdot F_{\mathrm{sig}}\left(\Delta m,\tau,\delta\tau\right) + \left(1-f_{\mathrm{sig}}\right)\cdot F_{\mathrm{bkgd}}\left(\Delta m,\tau,\delta\tau\right)\,,\nonumber
\label{Eqn:FullPDF}
\end{equation}

\noindent where $f_{\mathrm{sig}}$ is the fraction of $\chic$ signal candidates determined by the fit, while $F_{\mathrm{sig}}$ and $F_{\mathrm{bkgd}}$ are pdfs that respectively model the signal and background components of the mass difference and pseudo-proper decay time distributions. The quantity $\delta\tau$ is the per-candidate uncertainty on the pseudo-proper decay time calculated from the covariance matrix of the vertex fit to the dimuon tracks. The distributions of the pseudo-proper decay time uncertainty $\delta\tau$ observed in the background ($\Delta m < 0.3\GeV$ or $\Delta m > 0.48\GeV$) and signal (background subtracted within $0.3 <\Delta m < 0.48\GeV$) regions are found to be consistent within their uncertainties. Consequently, terms for the distribution of $\delta\tau$ factorise in the likelihood and are not included~\cite{Punzi}.

The signal pdf $F_{\mathrm{sig}}$ is composed of several components,

\begin{align*}
F_{\mathrm{sig}}\left(\Delta m,\tau,\delta\tau\right) = & f^{\mathrm{P}}_{\mathrm{sig}}\cdot\left[ f^{\mathrm{P}}_{0}\cdot M_{0}\left(\Delta m\right)+ \left(1-f^{\mathrm{P}}_{0}\right)\cdot\left(f^{\mathrm{P}}_{1}\cdot M_{1}\left(\Delta m\right) \right.\right.\nonumber \\
&\left.\left. +\left(1-f^{\mathrm{P}}_{1}\right)\cdot M_{2}\left(\Delta m\right)\right)\right]\cdot T^{\mathrm{P}}_{\mathrm{sig}}\left(\tau,\delta\tau\right)\nonumber \\
&+\left(1-f^{\mathrm{P}}_{\mathrm{sig}}\right)\cdot\left[ f^{\mathrm{NP}}_{0}\cdot M_{0}\left(\Delta m\right) + \left(1-f^{\mathrm{NP}}_{0}\right)\cdot\left(f^{\mathrm{NP}}_{1}\cdot M_{1}\left(\Delta m\right) \right.\right.\nonumber\\
&\left.\left. +\left(1-f^{\mathrm{NP}}_{1}\right)\cdot M_{2}\left(\Delta m\right)\right)\right]\cdot T^{\mathrm{NP}}_{\mathrm{sig}}\left(\tau,\delta\tau\right)\,, \nonumber\\
\end{align*}

\noindent where the pdfs $M_{J}\left(\Delta m\right)$ model the $\chicj J$ signal components of the $\Delta m$ distribution and $T_{\mathrm{sig}}^{\mathrm{(N)P}}$ model the (non-)prompt $\chic$ signal contributions to the pseudo-proper decay time distribution. 
The parameter $f^{\mathrm{P}}_{\mathrm{sig}}$ is the fraction of the total $\chic$ signal that is promptly produced, $f^{\mathrm{(N)P}}_{0}$ is the fraction of (non-)prompt signal identified as $\chicj0$ and $f^{\mathrm{(N)P}}_{1}$ is the fraction of (non-)prompt signal (excluding $\chicj0$ contributions) identified as $\chicj1$.

The $\chicj1$ and $\chicj2$ signal pdfs $M_{1,2}\left(\Delta m\right)$ are each described by a Crystal Ball (CB) function~\cite{CBFunction}. The CB function is characterised by a Gaussian core with a width $\sigma_{i}$ (associated with the pdf $M_{i}\left(\Delta m\right)$) and a power law low-mass tail described by the parameter $n$. The transition between the core and tail is described by the threshold parameter $\alpha$. The mass resolutions of prompt and non-prompt $\chic$ are found to be consistent from MC simulation studies. The natural widths of the $\chicj1$ and $\chicj2$ states are sufficiently small (below $2\MeV$~\cite{PDG}) relative to the detector mass resolution (around $10\MeV$) that their effects can be neglected. The peak positions of both the $\chicj1$ and $\chicj2$ CB functions are fixed to the world average values of their respective masses~\cite{PDG} multiplied by a common scale factor $\kappa$. The factor $\kappa$ is a free parameter in the fit and is present to account for electron energy losses that result in a small momentum scale shift. The fitted values are typically around $\kappa=0.98$. The two parameters $\alpha$ and $n$ of the CB functions are found to be consistent for both states and are fixed to values determined from fits to MC simulation samples. The resolution parameter $\sigma_{1}$ is determined by the fit with the constraint $\sigma_{2} = 1.07\times\sigma_{1}$ applied, where the value of the scale factor is extracted from MC simulation studies. The $\chicj0$ signal pdf, $M_{0}\left(\Delta m\right)$, is modelled by the sum of a CB function and a Gaussian pdf; all parameters describing the shape of the $\chicj0$ signal are fixed to values determined from MC simulation (including a natural width of $(10.3\pm0.6)\MeV$~\cite{PDG}), while the peak position of the $\chicj0$ signal is a free parameter in the fit. The $\chicj0$ peak is clearly visible in the $\Delta m$ distribution shown in figure~\ref{Fig:FullFit}. However, the $\chicj0$ yield is not reported because the statistical significance of the signal is marginal when the data are fitted in separate bins of $\ptjpsi$ and $\ptchic$.

The pdfs $T^{\mathrm{P}}_{\mathrm{sig}}$ and $T^{\mathrm{NP}}_{\mathrm{sig}}$ describing the prompt and non-prompt $\chic$ signal contributions to the pseudo-proper decay time distributions are modelled by a delta function $\delta\left(\tau\right)$ and an exponential function $\exp{\left(-\tau/\tau_{\mathrm{sig}}\right)}$, where $\tau_{\mathrm{sig}}$ is a free parameter in the fit. Both $\chic$ signal pseudo-proper decay time pdfs are convolved with the function $R\left(\tau^{\prime}-\tau,\delta\tau\right)$ describing the experimental resolution in pseudo-proper decay time. The resolution function $R$ is described by a Gaussian function with a mean value of zero and width $S\cdot\delta\tau$ where $S$ is a scale factor that is determined by the fit, while $\delta\tau$ is the per-candidate uncertainty on the pseudo-proper decay time $\tau$.

The background pdf $F_{\mathrm{bkgd}}$ is composed of two components,

\begin{align*}
F_{\mathrm{bkgd}}\left(\Delta m,\tau,\delta\tau\right) = & f^{\mathrm{P}}_{\mathrm{bkgd}}\cdot M^{\mathrm{P}}_{\mathrm{bkgd}}\left(\Delta m\right)\cdot T^{\mathrm{P}}_{\mathrm{bkgd}}\left(\tau,\delta\tau\right) \nonumber \\
&+ \left(1-f^{\mathrm{P}}_{\mathrm{bkgd}}\right)\cdot M^{\mathrm{NP}}_{\mathrm{bkgd}}\left(\Delta m\right)\cdot T^{\mathrm{NP}}_{\mathrm{bkgd}}\left(\tau,\delta\tau\right)\,, \nonumber \\
\end{align*}

\noindent where $M^{\mathrm{(N)P}}_{\mathrm{bkgd}}$ describes the (non-)prompt background contributions to the $\Delta m$ distribution and $T^{\mathrm{(N)P}}_{\mathrm{bkgd}}$ describes the (non-)prompt background contributions to the pseudo-proper decay time distribution. The parameter $f^{\mathrm{P}}_{\mathrm{bkgd}}$ is the fraction of the background that is promptly produced. The functions $M^{\mathrm{(N)P}}_{\mathrm{bkgd}}$ are both modelled by the function

\begin{equation}
M_{\mathrm{bkgd}}\left(\Delta m\right) = \mathrm{erf}\left(A\cdot\left(\Delta m - m_{0}\right)\right)\cdot\exp\left(B\cdot\left(\Delta m - m_{0}\right)\right) + C\cdot\left(\Delta m - m_{0}\right)^{2}\,,\nonumber
\label{Eqn:MassBkgdPDF}
\end{equation}

\noindent where all four parameters ($A$, $B$, $C$ and $m_{0}$) in the pdf are determined by the fit. The shape of the background pdf is motivated by studies with MC simulation samples and a sample of $\mumu\gamma$ candidates in data with dimuon invariant masses in the sidebands of the $\Jpsi$ peak. The prompt and non-prompt background pdfs each use an independent set of four parameters. Endowing both background pdfs with an independent set of parameters is motivated by the observation that the shape of the background contribution to the $\Delta m$ distribution in data varies significantly as a function of pseudo-proper decay time. The pdf $T^{\mathrm{P}}_{\mathrm{bkgd}}\left(\tau,\delta\tau\right)$ is modelled with a delta function convolved with the pseudo-proper decay time resolution function $R$. The pdf $T^{\mathrm{NP}}_{\mathrm{bkgd}}\left(\tau,\delta\tau\right)$ consists of two components,

\begin{align*}
T^{\mathrm{NP}}_{\mathrm{bkgd}}\left(\tau,\delta\tau\right) =& \left[\frac{g_{\mathrm{bkgd}}}{\tau_{\mathrm{bkgd}}}\cdot\exp\left(-\tau^{\prime}/\tau_{\mathrm{bkgd}}\right) \right.\nonumber\\ 
&\left.+ \frac{(1-g_{\mathrm{bkgd}})}{2\tau_{\mathrm{sym}}}\cdot\exp\left(-|\tau^{\prime}|/\tau_{\mathrm{sym}}\right)\right]\otimes R\left(\tau^{\prime}-\tau,\delta\tau\right) \,, \nonumber\\
\end{align*}

\noindent where $g_{\mathrm{bkgd}}$ determines the relative mixture of the single- and double-sided exponential components. The parameters $\tau_{\mathrm{bkgd}}$ and $\tau_{\mathrm{sym}}$ determine the shapes of the single- and double-sided exponential components, respectively. The result of the fit described above to the inclusive data sample with $10 \leq \ptjpsi < 30\GeV$ is shown by the projections onto the mass difference and pseudo-proper decay time axes as shown in figure~\ref{Fig:FullFit}. The purity of the selected $\Jpsi$ candidates is around $90\%$, with no strong dependence on pseudo-proper decay time. A larger background contribution to the $m\left(\mumu\gamma\right) - m\left(\mumu\right)$ distribution, relative to the $\chic$ signal, is observed for $\mumu\gamma$ candidates with longer pseudo-proper decay times. This behaviour is consistent with the expectation from MC simulation and is due to the presence of additional charged particles and photons produced close to the dimuon system in the decays of $b$-hadrons.

Bin migrations in the measured $\ptchic$ and $\ptjpsi$ distributions are corrected with the method described in refs~\cite{ATLAS_Jpsi,ATLAS_Upsilon}. The approach involves fitting the measured $\ptchic$ and $\ptjpsi$ distributions with a smooth analytic function that is convolved with a resolution function determined from MC simulation. The ratio of the functions with and without convolution is used to deduce a correction factor for each measured bin in $\ptchic$ and $\ptjpsi$. The average corrections for the cross-sections measured as a function of $\ptjpsi$ is $0.5\%$. Corrections to the cross-sections measured as a function of $\ptchic$ are significantly larger, around $4\%$ on average, due to an asymmetric experimental resolution in $\ptchic$ caused by electron energy loss through bremsstrahlung.

\begin{figure}

\begin{center}
\includegraphics[width=0.7\textwidth]{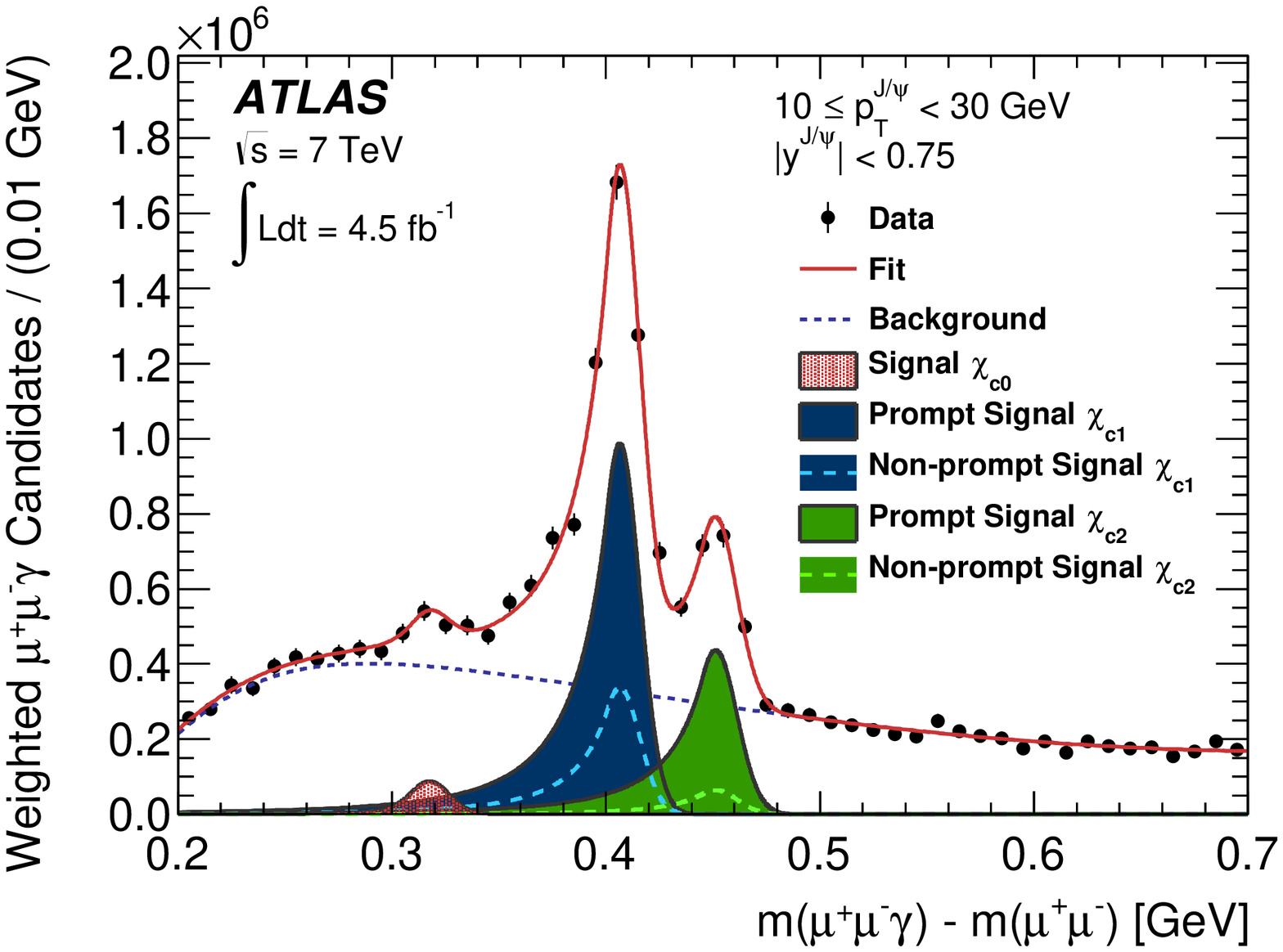}

\includegraphics[width=0.7\textwidth]{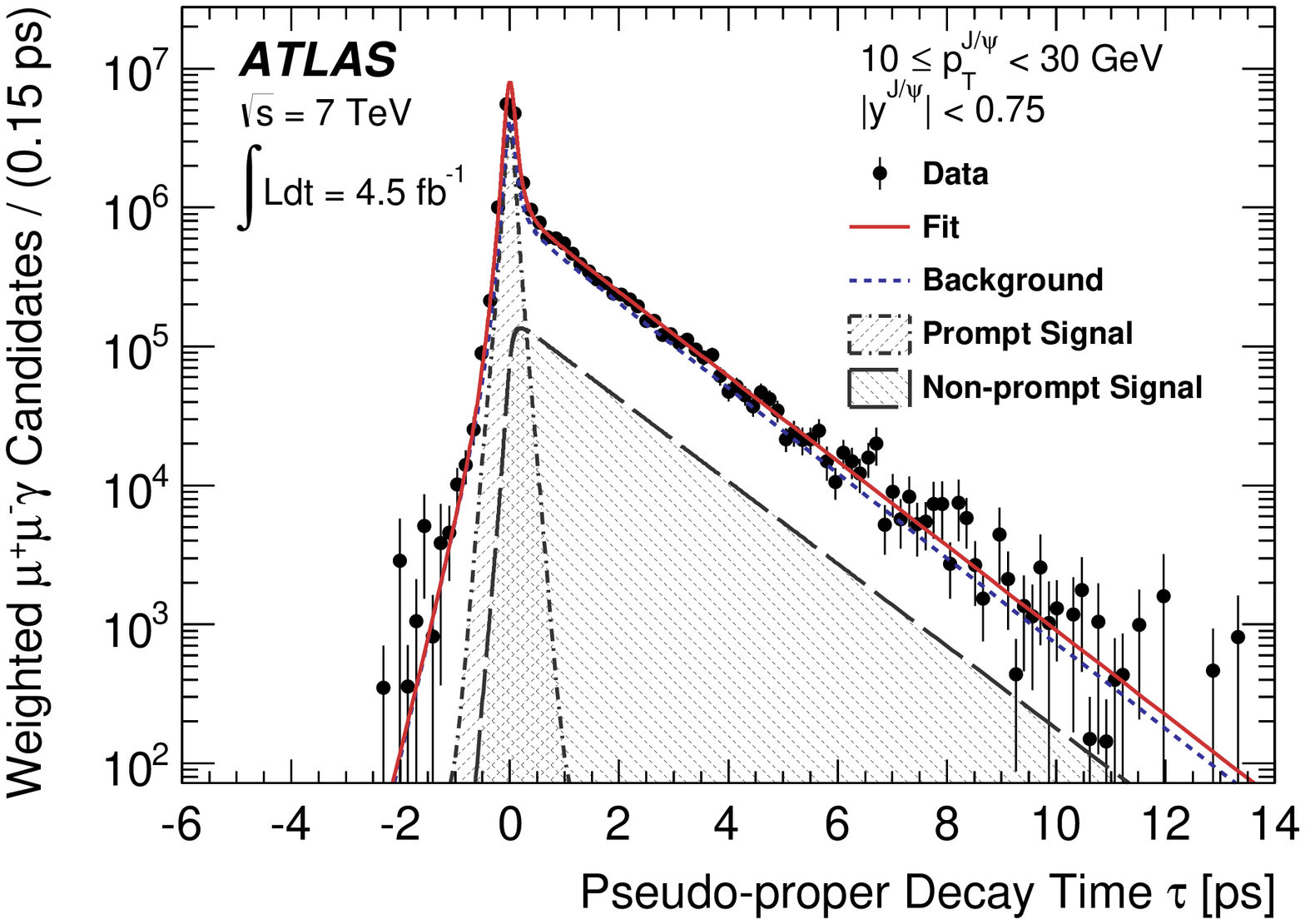}
\end{center}

\caption{The mass difference $\Delta m = m\left(\mumu\gamma\right) - m\left(\mumu\right)$ distribution (top) for $\chic$ candidates reconstructed within $10 \leq \ptjpsi < 30\GeV$ and $|\yjpsi| < 0.75$. The pseudo-proper decay time $\tau$ distribution of the same sample of $\chic$ candidates is also shown (bottom). Both distributions are corrected for acceptance and experimental efficiency (the prompt $\chicj1$ acceptance correction is shown here for demonstration). The result of the simultaneous fit to both distributions is shown by the overlaid solid red lines. The fitted $\chicj J$ signals are shown by the shaded regions while the fitted background distributions are shown by the dashed blue lines.}
\label{Fig:FullFit}
\end{figure}

\clearpage


\section{Systematic uncertainties}

The relevant sources of systematic uncertainty on the prompt and non-prompt $\chicj1$ and $\chicj2$ cross-section measurements have been identified and their effects are discussed and quantified below.

\begin{description}

\item[\textbf{Statistical uncertainties in efficiency corrections}] \hfill \\
One component of the systematic uncertainty in the efficiency corrections is statistical in nature due to the finite size of the MC and data control samples used to derive the efficiencies. The corresponding uncertainty in the $\chic$ yields is quantified by repeatedly varying the values of the efficiencies within their statistical uncertainties by a random amount and recalculating the weights used to extract the corrected $\chic$ yields. The systematic uncertainty is estimated from the RMS of the distribution of the average weight in each $\ptchic$ and $\ptjpsi$ bin.

\item[\textbf{Conversion probability}] \hfill \\
The conversion probability $P_{\mathrm{conv}}$ derived using MC simulation samples is sensitive to the modelling of the ID material distribution in the ATLAS detector simulation. As discussed in section~\ref{Sec:ConvEff}, the description of the ID material distribution in MC simulation is found to agree with the distribution observed in data~\cite{Material}. The uncertainties in the modelling of the mass distribution are used to define an alternative ATLAS detector model with a larger amount of material in the conversion fiducial region. Samples of $\chic\to\Jpsi\,\gamma$ events are generated with both the nominal and the alternative detector model, using the same method as described in ref.~\cite{Higgs}. The difference in the simulated conversion probabilities is taken as an estimate of the corresponding systematic uncertainty.

\item[\textbf{Converted-photon reconstruction efficiency}] \hfill \\
The converted-photon reconstruction efficiency $\epsilon_{\mathrm{conv}}$ derived using MC simulation samples is sensitive to any potential differences in the behaviour of the photon conversion reconstruction algorithm in data and simulation. These effects are studied by comparing several sensitive distributions (including conversion vertex position, vertex fit quality and $\ee$ track hits) of a sample of photon conversions in data and simulation. The systematic uncertainty in $\epsilon_{\mathrm{conv}}$ due to residual simulation mis-modelling effects is estimated to be $\pm9\%$. The systematic uncertainty on the determination of $\epsilon_{\mathrm{conv}}$ due to potential residual mismodelling of bin migrations in $\ptgamma$ is estimated to be $\pm5\%$ from several self consistency tests performed with MC simulation samples.

\item[\textbf{Acceptance}] \hfill \\
The per-candidate acceptance correction for the measurements binned in $\ptchic$ is calculated from the reconstructed value of $\ptchic$ for each $\chic$ candidate. This reconstructed value of $\ptchic$ is scaled by around $+0.5\%$ to avoid an over-correction due to the asymmetric experimental resolution in $\ptchic$. The correction procedure is verified with simulation studies and has a systematic uncertainty of $\pm2\%$. The acceptance corrections for the measurements binned in $\ptjpsi$ are sensitive to the $\ptchic$ distribution used as an input to the acceptance simulation. This sensitivity is studied by varying the analytic function used to fit the measured $\ptchic$ distribution. The systematic uncertainty in the acceptance corrections due to the fitted parametrisation of the measured $\ptchic$ distributions is estimated to be between $4\mhyphen8\%$, depending on the individual $\chicj J$ yield. The systematic uncertainty in the acceptance correction, due to the unknown $\chic$ spin-alignment, is evaluated by comparing the acceptance calculated assuming isotropic decay angular distributions to that calculated with the angular distributions corresponding to the spin-alignment scenarios shown in Table~\ref{Table:Pol}. This comparison is used to derive an uncertainty envelope associated with the unknown $\chic$ spin-alignment. The spin-alignment envelope is treated as a separate uncertainty and is discussed in more detail in section~\ref{Sec:Results}.

\item[\textbf{Fit model}] \hfill \\
The systematic uncertainty on the $\chicj J$ yields due to the fit model is quantified with a MC pseudo-experiment approach. Pseudo-data samples are generated from the nominal fit result in individual bins of $\ptchic$ and $\ptjpsi$. Each pseudo-data sample is fitted with the nominal fit model and five alternative fit models that include a variety of alterations as discussed below.

\begin{itemize}
\item Each of the fixed $\alpha$ and $n$ parameters in the $\chicj1$ and $\chicj2$ signal CB functions is individually released to be  determined by the fit.
\item The scaling of the $\chicj1$ and $\chicj2$ $\Delta m$ resolution parameters $\sigma_{1}$ and $\sigma_{2}$ is removed and both parameters are independently determined by the fit.
\item The fit is repeated with the $\chicj0$ signal component of the pdf removed. This test is motivated by the fact that the $\chicj0$ signal is insignificant in some of the $\pt$ bins studied.
\item An alternative background pdf (with four free parameters) for the mass difference distribution is tested.
\end{itemize}

The systematic uncertainty due to the fit model is then estimated from the mean of the distribution of the relative changes in the yield between the nominal and alternative models. The systematic uncertainty due to the fit model for the prompt and non-prompt cross-section ratios and non-prompt fractions is evaluated in the same way to ensure correlations between the signal components in the evaluation of the systematic uncertainty are taken into account.

\item[\textbf{Integrated luminosity}]  \hfill \\
The uncertainty in the integrated luminosity of the data sample used is estimated to be $1.8\%$. The methods used to determine this uncertainty are described in detail in ref.~\cite{Lumi}. This systematic uncertainty does not affect the cross-section ratios or non-prompt fractions.  
\end{description}

\noindent The individual sources of systematic uncertainty are summarised in tables~\ref{Tab:ErrorsJpsiPt} and~\ref{Tab:ErrorsChiPt}. The largest sources of systematic uncertainty are associated with the photon conversion reconstruction efficiency, the fit model and the acceptance. The individual sources of systematic uncertainty studied are not strongly correlated and the total systematic uncertainty on the measurements is obtained by adding the individual systematic uncertainties in quadrature. These uncertainties are generally much smaller than the uncertainty associated with the unknown $\chic$ spin-alignment. 

\clearpage

\begin{table}[h]
\begin{center}
\begin{tabular}{| l | C | C | C | C |}
\hline
\multirow{3}{*}{Binning: $\ptjpsi$}& \multicolumn{4}{c|}{Fractional Uncertainty [$\%$]} \\
\cline{2-5}
&\multicolumn{2}{c|}{Prompt}&\multicolumn{2}{c|}{Non-prompt}\\
\cline{2-5}
& $\chicj1$ & $\chicj2$ & $\chicj1$ & $\chicj2$ \\
\hline
Muon reco. efficiency     & 1 & 1 & 1 & 1 \\
Trigger efficiency      & 4 & 4 & 4 & 4 \\
Converted-photon reco. efficiency & 11 & 11 & 11 & 11 \\
Conversion probability    & 4 & 4 & 4 & 4 \\
Acceptance          & 4 & 4 & 5 & 8 \\
Fit model           & 2 & 3 & 3 & 9 \\
\hline
Total systematic      & 13 & 13 & 13 & 17 \\
\hline
Spin-alignment envelope (upper) & 34 & 36 & 32 & 36 \\
Spin-alignment envelope (lower) & 13 & 23 & 13 & 23 \\
\hline
\end{tabular}
\end{center}
\caption{The individual contributions to the systematic uncertainty on the cross-section measurements binned in $\ptjpsi$, averaged across all $\ptjpsi$ bins. The common contributions of integrated luminosity ($1.8\%$) and track reconstruction ($1\%$) are not shown. The average variation in the cross-sections due to the envelope of all possible spin-alignment scenarios is also shown (``upper'' denotes the positive variation while ``lower'' denotes the negative variation).}
\label{Tab:ErrorsJpsiPt}
\end{table}

\begin{table}[h]
\begin{center}
\begin{tabular}{| l | C | C | C | C |}
\hline
\multirow{3}{*}{Binning: $\ptchic$}& \multicolumn{4}{c|}{Fractional Uncertainty [$\%$]} \\
\cline{2-5}
\multirow{2}{*}{}& \multicolumn{2}{c|}{Prompt} & \multicolumn{2}{c|}{Non-prompt} \\
\cline{2-5}
& $\chicj1$ & $\chicj2$ & $\chicj1$ & $\chicj2$ \\
\hline
Muon reco. efficiency     & 1 & 1 & 1 & 1 \\
Trigger efficiency      & 3 & 4 & 4 & 4 \\
Converted-photon reco. efficiency & 11 & 11 & 11 & 11  \\
Conversion probability    & 4 & 4 & 4 & 4 \\
Acceptance          & 2 & 2 & 2 & 2 \\
Fit model           & 2 & 3 & 3 & 8  \\
\hline
Total systematic      & 12 & 12 & 12 & 14 \\
\hline
Spin-alignment envelope (upper) & 29 & 31 & 29 & 31 \\
Spin-alignment envelope (lower) & 11 & 20 & 11 & 20 \\
\hline
\end{tabular}
\end{center}
\caption{The individual contributions to the systematic uncertainty on the cross-section measurements binned in $\ptchic$, averaged across all $\ptchic$ bins. The common contributions of integrated luminosity ($1.8\%$) and track reconstruction ($1\%$) are not shown. The average variation in the cross-sections due to the envelope of all possible spin-alignment scenarios is also shown (``upper'' denotes the positive variation while ``lower'' denotes the negative variation).}
\label{Tab:ErrorsChiPt}
\end{table}

\clearpage


\section{Results and interpretation}
\label{Sec:Results}

The differential cross-sections of prompt and non-prompt $\chicj1$ and $\chicj2$ production are measured in bins of $\ptjpsi$ and $\ptchic$ within the rapidity region $|\yjpsi|<0.75$. The results measured as a function of $\ptjpsi$ and $\ptchic$ are presented within the regions $10\leq\ptjpsi<30\GeV$ and $12\leq\ptchic<30\GeV$, respectively. The measurements of the prompt production of $\chicj1$ and $\chicj2$ as a function of $\ptjpsi$ are combined with existing measurements~\cite{ATLAS_Jpsi} of prompt $\Jpsi$ production to determine the fraction of prompt $\Jpsi$ produced in $\chic$ decays. The production rate of $\chicj2$ relative to $\chicj1$ is measured for prompt and non-prompt $\chic$ as a function of $\ptjpsi$. The fractions of $\chicj1$ and $\chicj2$ produced in the decays of $b$-hadrons are also presented as a function of $\ptchic$. While the prompt and non-prompt $\chicj1$ and $\chicj2$ yields are necessarily extracted from separate fits, statistical correlations between the yields are taken into account in the calculation of the statistical uncertainties on the cross-section ratios and non-prompt fractions. Tabulated results for all of the measurements are included in appendix~\ref{Sec:Appendix}.

The spin-alignment of the $\chic$ mesons produced at the LHC is unknown. All measurements are corrected for detector acceptance assuming isotropic angular distributions for the decay of the $\chic$ states. The total uncertainty in the measurements due to this assumption is obtained by comparing the acceptance calculated with the extreme $\chic$ spin-alignment scenarios discussed in section~\ref{Sec:Acc} to the central value, calculated with the isotropic scenario. The maximum uncertainty due to spin-alignment effects averaged over each measurement bin is shown in tables~\ref{Tab:ErrorsJpsiPt} and~\ref{Tab:ErrorsChiPt} and for each measurement bin in the tabulated results in appendix~\ref{Sec:Appendix}. The uncertainty due to spin-alignment effects is not included in the total systematic uncertainty. Factors that can be used to scale the central cross-section values to any of the individual spin-alignment scenarios studied are also provided in table~\ref{Table:ScaleFactors} of appendix~\ref{Sec:Appendix}.

\subsection{Differential cross-sections}

Differential cross-sections for prompt $\chicj1$ and $\chicj2$ production are measured as a function of both $\ptjpsi$ and $\ptchic$ within the region $|\yjpsi|<0.75$. These results are corrected for acceptance assuming isotropic decay angular distributions for the $\chicj1$ and $\chicj2$ states and are shown in figures~\ref{Fig:CSPrompt_JpsiPt} and~\ref{Fig:CSPrompt_ChiPt}. The position along the $\pt$ axis of each of the data points shown in these figures is adjusted to reflect the average value of the transverse momentum, $\langle\ptjpsi\rangle$ or $\langle\ptchic\rangle$, for the $\chic$ candidates within that $\pt$ bin, after all acceptance and efficiency corrections have been applied. The measurements are compared with the predictions of next-to-leading-order (NLO) non-relativistic QCD (NRQCD)~\cite{NRQCD,HELAC,JpsiNRQCD}, the $k_{\mathrm{T}}$ factorisation approach~\cite{kT1,kT2} and leading-order (LO) colour-singlet model (CSM)~\cite{CHIGEN} calculations. The NRQCD factorisation approach separates the perturbative production of a heavy quark pair (in a colour-singlet or -octet state) from the non-perturbative evolution of a heavy quark pair into a quarkonium state~\cite{BBL}. The long-distance effects are described by matrix elements that are determined by fitting experimental data~\cite{QWG}. For the predictions shown, the NRQCD long-distance matrix elements are extracted from measurements of $\Jpsi$ and $\psi(2S)$ production at the Tevatron as described in ref.~\cite{JpsiNRQCD}. In the CSM, heavy quark pairs are produced directly in a colour-singlet state (described by perturbative QCD) and a potential model is used to describe the formation of the bound state~\cite{CSM2,CSM3,CSM4,CSM5}. In the high $\pt$ region studied, $gg\to\chicj J g$ processes constitute the dominant contribution to the CSM prediction. The $k_{\mathrm{T}}$ factorisation approach convolves a partonic cross-section from the CSM with an un-integrated gluon distribution that depends on both longitudinal and transverse momentum (as opposed to the collinear approximation, which neglects parton transverse momentum) to calculate the hadronic cross-section. The shaded uncertainty bands of the NRQCD and CSM predictions are derived from factorisation and renormalisation scale uncertainties, and the NRQCD uncertainty also includes a contribution from the extraction of NRQCD long distance matrix elements from data. Good agreement between the NRQCD calculation and the measurements is observed. The $k_{\mathrm{T}}$ factorisation approach predicts a cross-section significantly in excess of the measurement while the LO CSM prediction significantly underestimates the data. This suggests that higher-order corrections or colour-octet contributions to the cross-sections not included in either prediction may be numerically important.

Differential cross-sections are also measured for non-prompt $\chicj1$ and $\chicj2$ production as functions of both $\ptjpsi$ and $\ptchic$ within the region $|\yjpsi|<0.75$, assuming isotropic decay angular distributions. The results are shown in figure~\ref{Fig:CSNonPrompt} and are compared to the fixed order next-to-leading-logarithm (FONLL) prediction for $b$-hadron production~\cite{FONLL,FONLL2}. These predictions are combined with measured momentum distributions of $\chicj1$ and $\chicj2$ in the $B^{\pm/0}$ rest frame for inclusive $B\to\chic X$ decays~\cite{BaBar_IncB}. This prediction is scaled assuming all $b$-quarks hadronise into $B^{\pm/0}$ mesons. The current world average values for the branching fractions ${\cal B}\left(B^{\pm/0}\to\chicj1 X\right) = (3.86\pm0.27)\times10^{-3}$ and ${\cal B}\left(B^{\pm/0}\to\chicj2 X\right) = (1.3\pm0.4)\times10^{-3}$ are used~\cite{PDG}. The shaded uncertainty band on the FONLL predictions represents the theoretical uncertainty due to factorisation and renormalisation scales, quark masses and parton distribution functions combined with the uncertainty on the branching fractions used to scale the predictions. The measurements generally agree with the FONLL predictions, though the data tend to lie slightly below the predictions at high $\pt$.

\clearpage

\begin{figure}
\begin{center}
\includegraphics[width=0.7\textwidth]{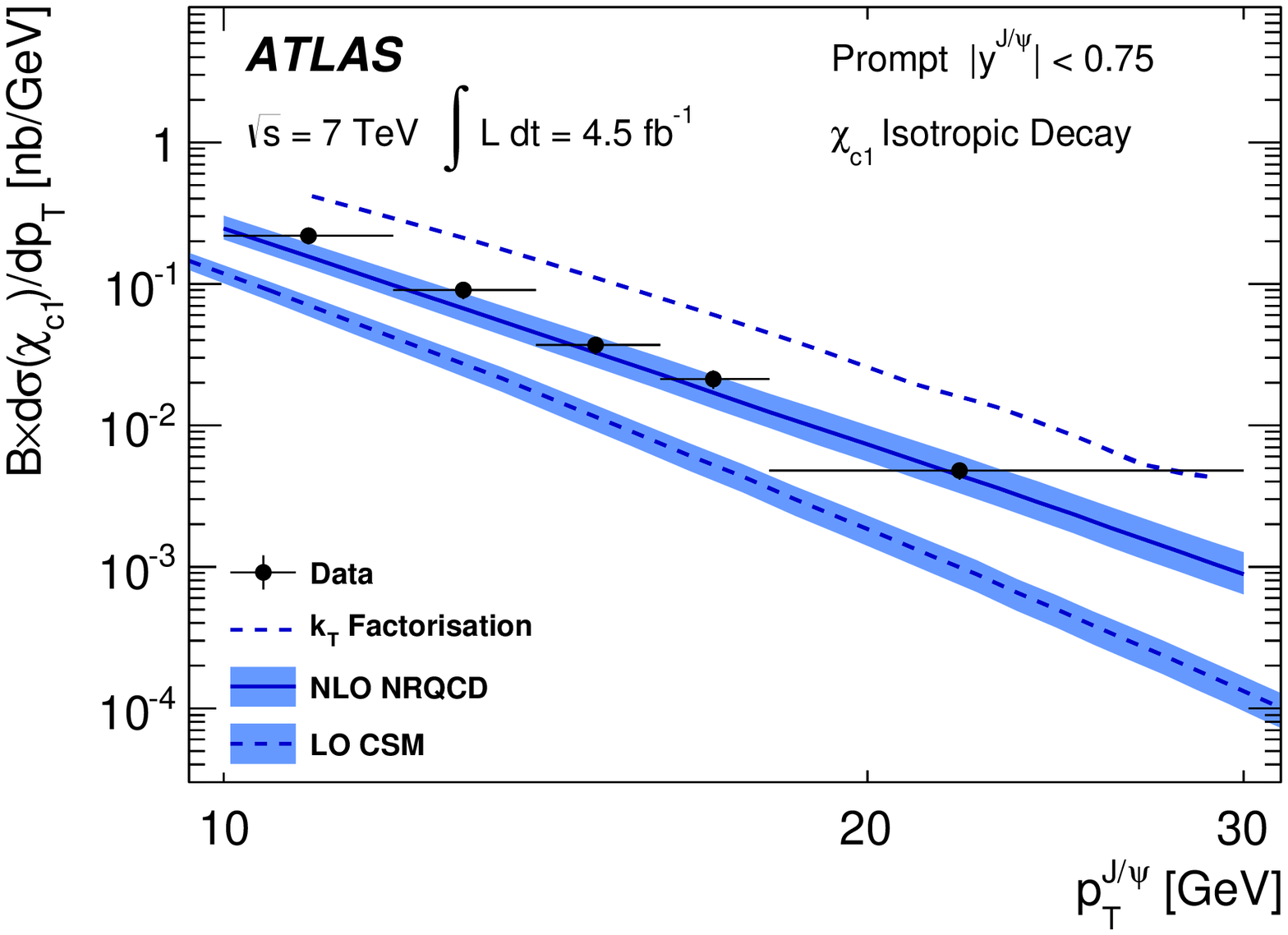}

\includegraphics[width=0.7\textwidth]{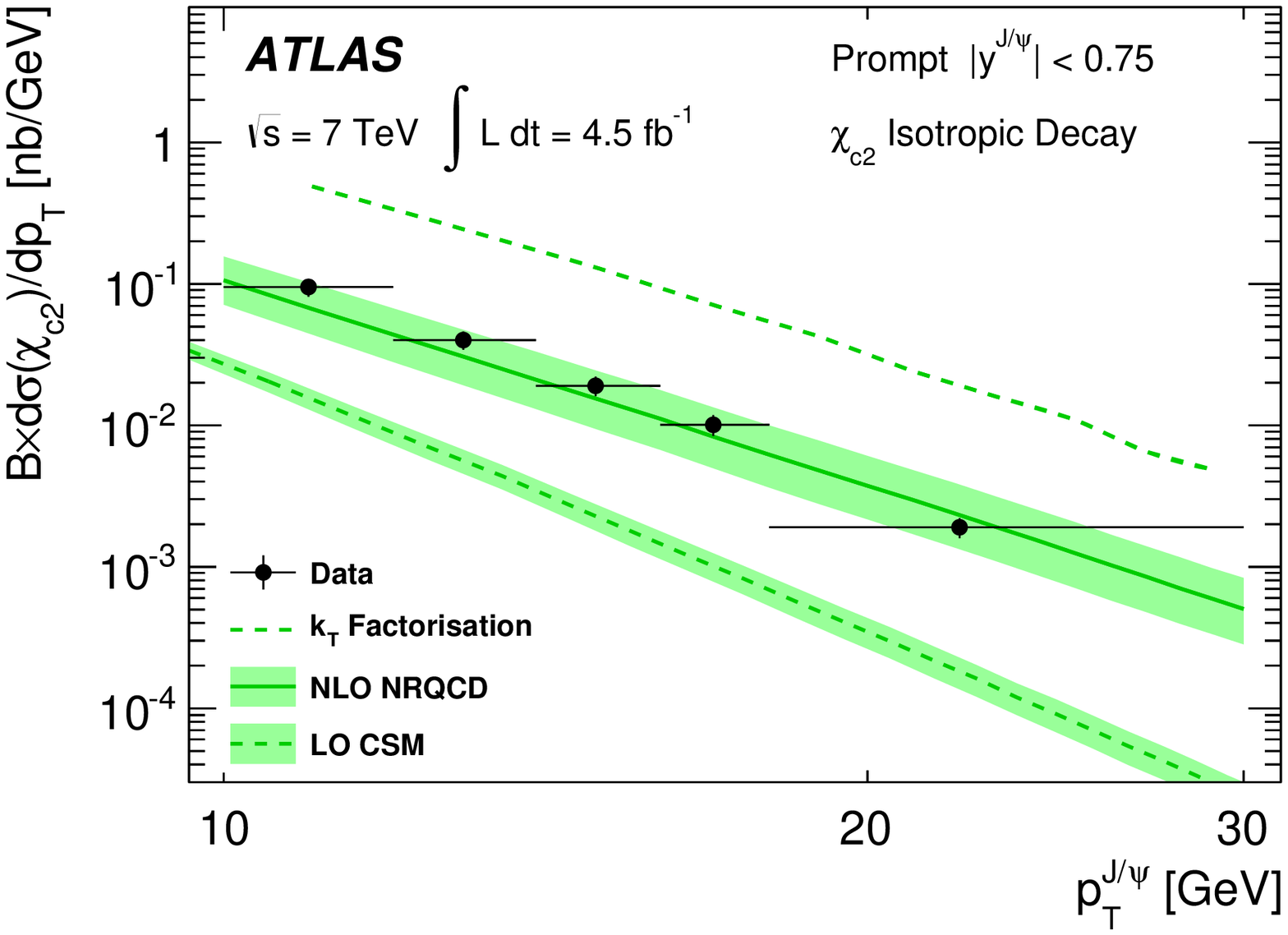}
\end{center}
\caption{Differential cross-sections for prompt $\chicj1$ (top) and $\chicj2$ (bottom) production as a function of $\ptjpsi$. The predictions of NLO NRQCD, the $k_{\mathrm{T}}$ factorisation model and the LO CSM are compared to the measurements. The positions of the data points within each bin reflect the average $\ptjpsi$ of the $\chic$ candidates within the bin. The error bars represent the total uncertainty on the measurement, assuming isotropic decay angular distributions (in some cases, the error bar is smaller than the data point). The factor $B$ denotes the product of branching fractions, $B={\cal B}\left(\chicj J \to\Jpsi\,\gamma\right)\cdot{\cal B}\left(\Jmumu\right)$.}
\label{Fig:CSPrompt_JpsiPt}
\end{figure}

\clearpage

\begin{figure}
\begin{center}
\includegraphics[width=0.7\textwidth]{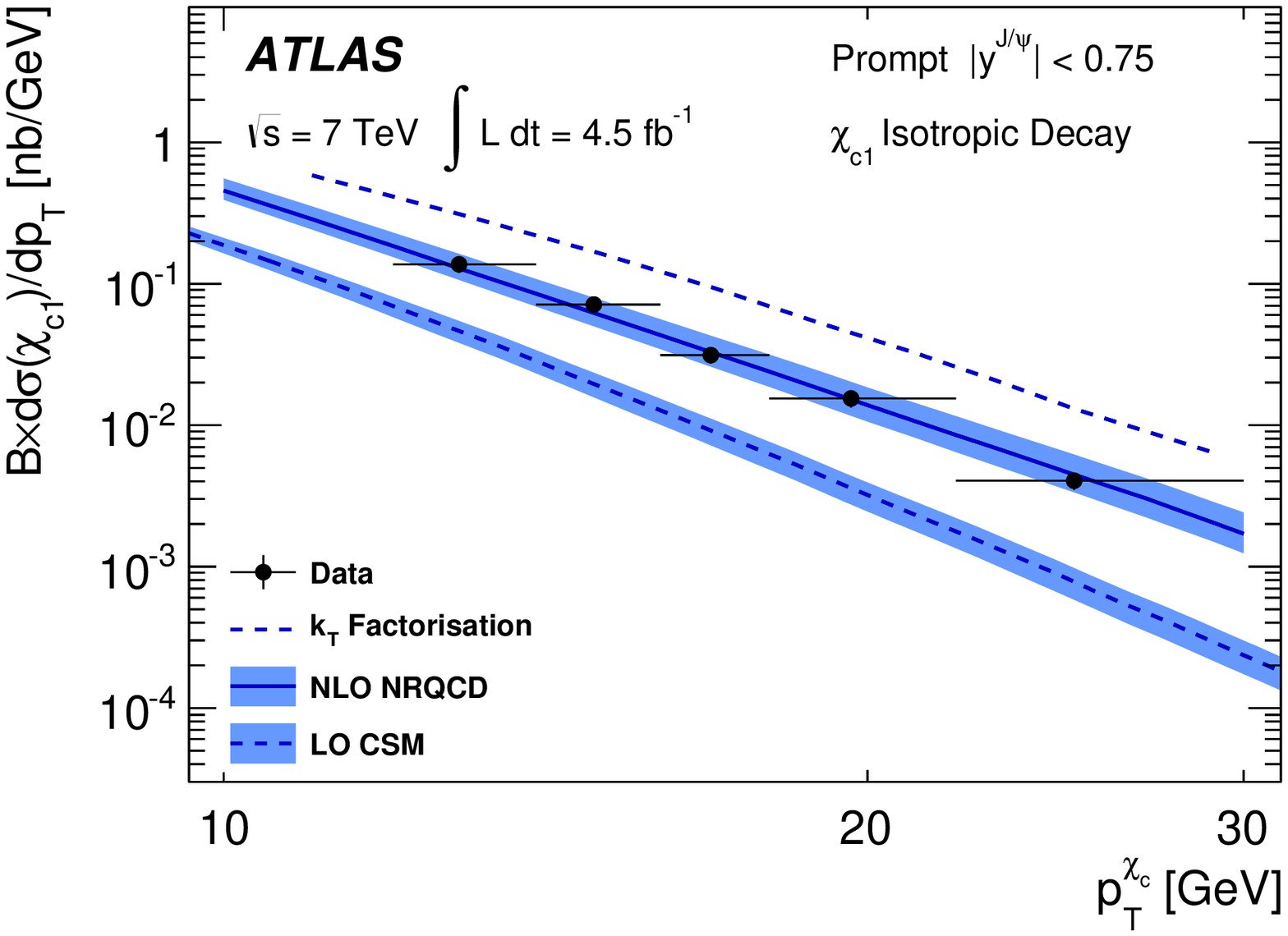}

\includegraphics[width=0.7\textwidth]{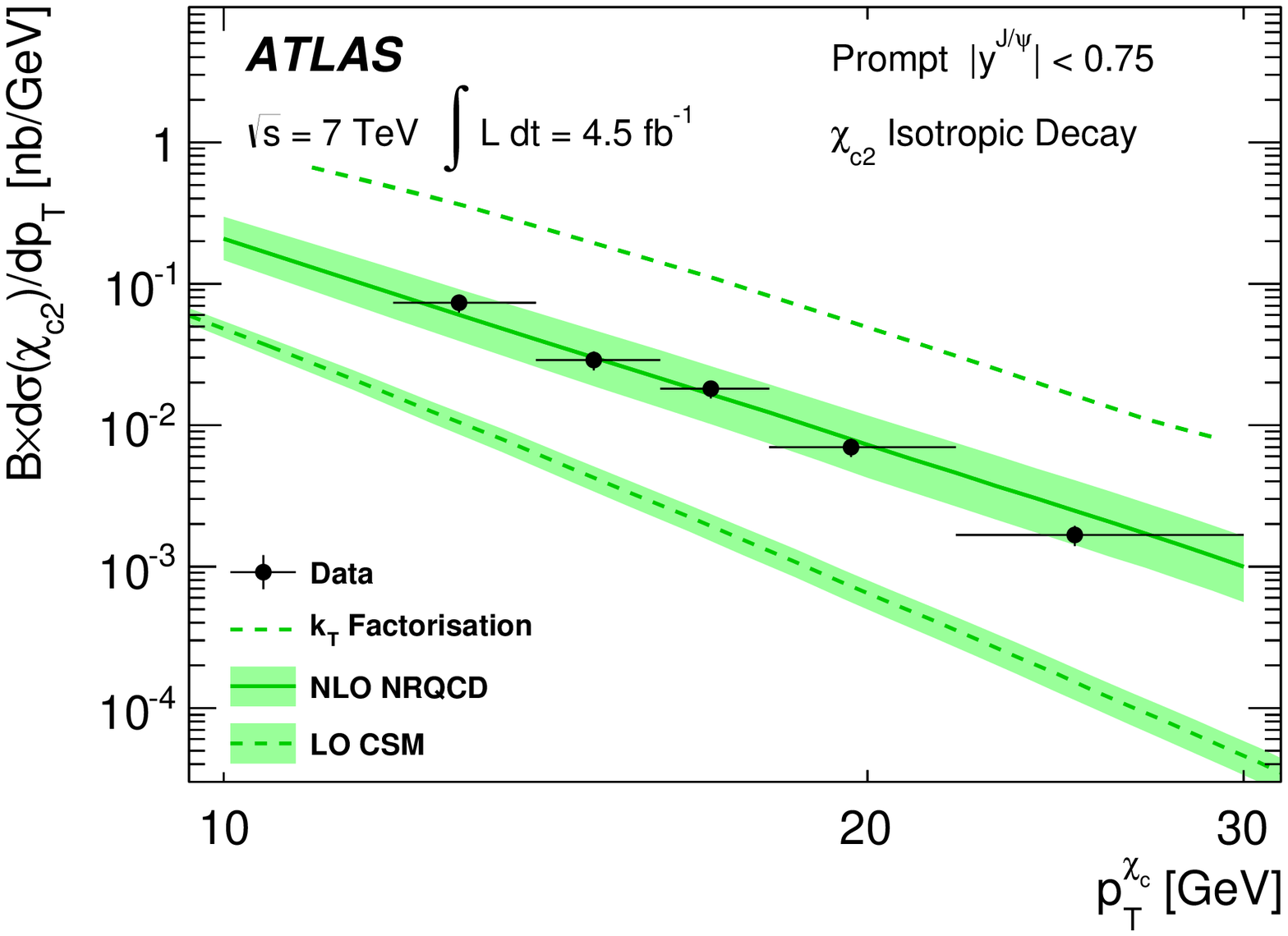}
\end{center}
\caption{Differential cross-sections for prompt $\chicj1$ (top) and $\chicj2$ (bottom) production as a function of $\ptchic$. The predictions of NLO NRQCD, the $k_{\mathrm{T}}$ factorisation model and the LO CSM are compared to the measurements. The positions of the data points within each bin reflect the average $\ptchic$ of the $\chic$ candidates within the bin. The error bars represent the total uncertainty on the measurement, assuming isotropic decay angular distributions (in some cases, the error bar is smaller than the data point). The factor $B$ denotes the product of branching fractions, $B={\cal B}\left(\chicj J \to\Jpsi\,\gamma\right)\cdot{\cal B}\left(\Jmumu\right)$.}
\label{Fig:CSPrompt_ChiPt}
\end{figure}

\clearpage

\begin{figure}
\begin{center}
\includegraphics[width=0.7\textwidth]{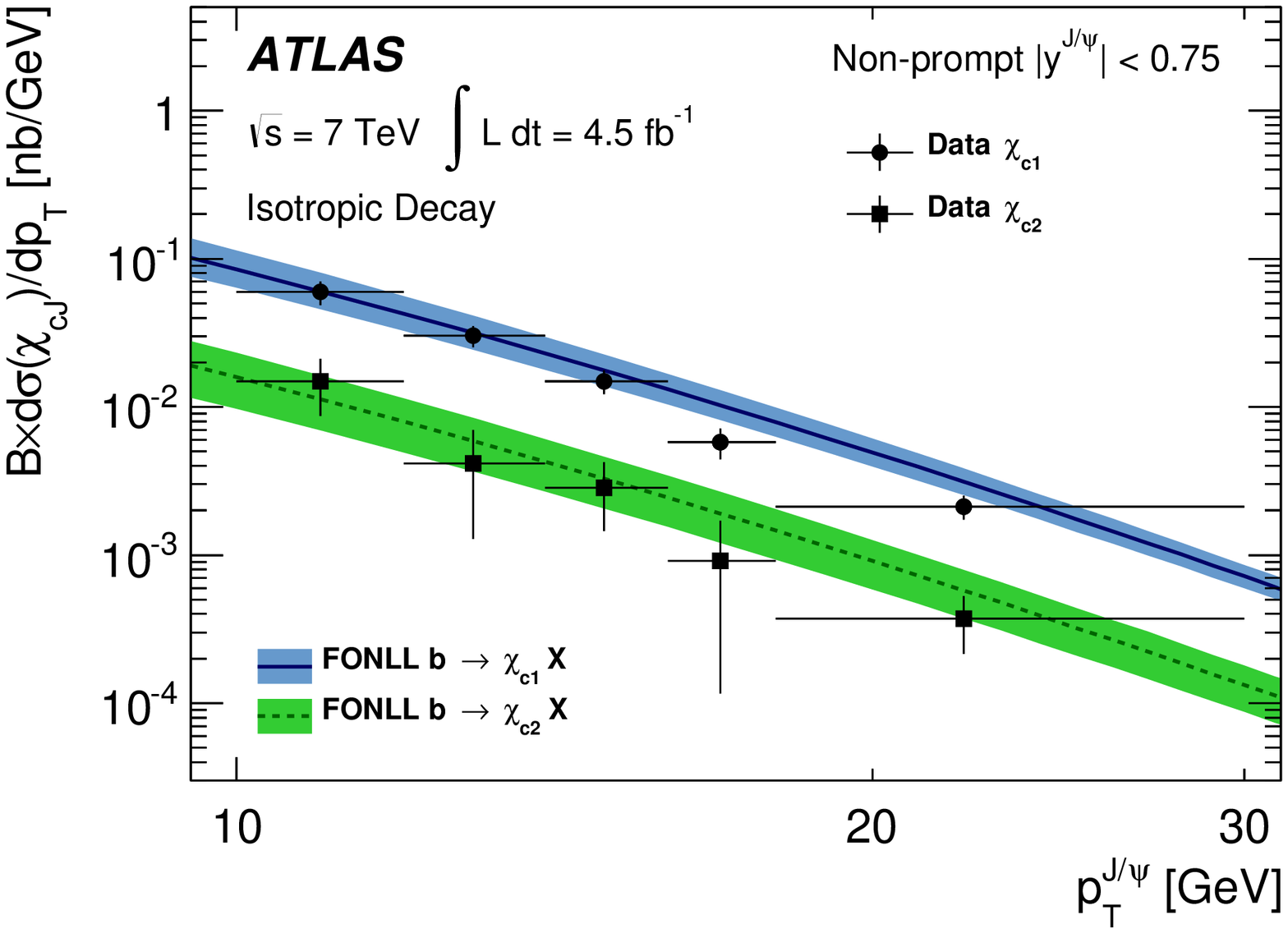}

\includegraphics[width=0.7\textwidth]{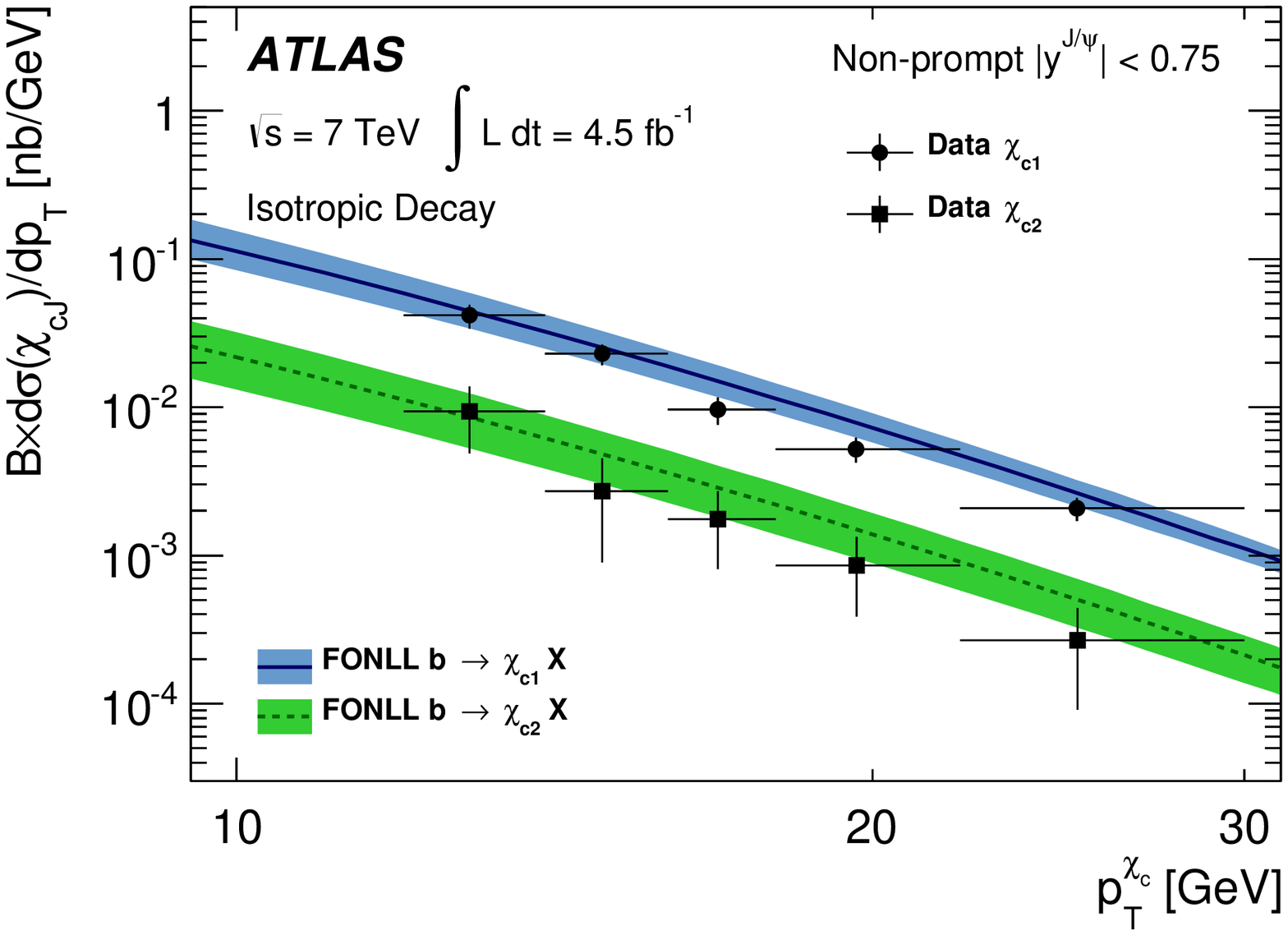}
\end{center}
\caption{Differential cross-sections for non-prompt $\chicj1$ and $\chicj2$ production as a function of $\ptjpsi$ (top) and $\ptchic$ (bottom). The predictions of FONLL are compared to the measurements. The positions of the data points within each bin reflect the average $\ptjpsi$ and $\ptchic$ of the $\chic$ candidates within the bin. The error bars represent the total uncertainty on the measurement, assuming isotropic decay angular distributions. The factor $B$ denotes the product of branching fractions, $B={\cal B}\left(\chicj J \to\Jpsi\,\gamma\right)\cdot{\cal B}\left(\Jmumu\right)$.}
\label{Fig:CSNonPrompt}
\end{figure}

\clearpage

\subsection{\boldmath Fraction of prompt $\Jpsi$ produced in $\chic$ decays}

The prompt $\chicj1$ and $\chicj2$ cross-sections are summed to provide their total contribution to the prompt $\Jpsi$ cross-section for each $\ptjpsi$ bin. The result is then divided by the prompt $\Jpsi$ cross-section measured by ATLAS~\cite{ATLAS_Jpsi} for each $\ptjpsi$ bin. The systematic uncertainties in the two measurements are treated as uncorrelated. This is motivated by the fact that the ATLAS measurement of the prompt $\Jpsi$ cross-section was performed with an independent data sample (recorded during the 2010 LHC run) and that the systematic uncertainties on the experimental efficiencies are evaluated using different methods and are all dominated by statistical effects. This provides a measurement of the fraction, $R_{\chic}$, of $\Jpsi$ produced in feed-down from $\chic$ decays, neglecting the small contribution from radiative $\chicj0$ decays, as a function of $\ptjpsi$. The measurements are shown in figure~\ref{Fig:JpsiRatio} and are compared to the predictions of NLO NRQCD and the LHCb measurement within the range $2.0 < \yjpsi < 4.5$~\cite{LHCB_ChiJpsiRatio}. The results show that between $20\%$ and $30\%$ of prompt $\Jpsi$ are produced in $\chic$ feed-down at high $\Jpsi$ transverse momentum. Motivated by recent measurements~\cite{CMS_JpsiPol,LHCb_JpsiPol}, the spin-alignment envelope for this fraction given in table~\ref{Tab:JpsiRatio} in appendix~\ref{Sec:Appendix} is calculated assuming no overall spin-alignment for promptly produced $\Jpsi$.

\begin{figure}
\begin{center}
\includegraphics[width=0.7\textwidth]{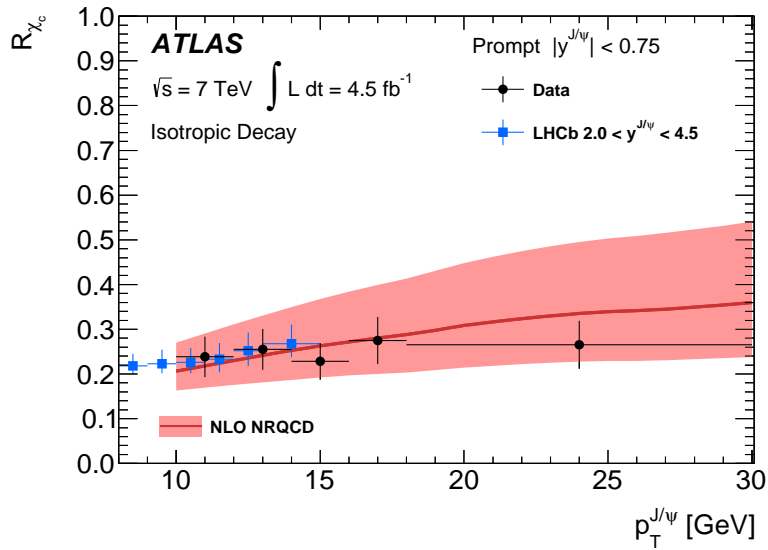}
\end{center}
\caption{The fraction, $R_{\chic}$, of prompt $\Jpsi$ produced in $\chic$ decays as a function of $\ptjpsi$. The measurements are compared to the prediction of NLO NRQCD. The measurement from LHCb~\cite{LHCB_ChiJpsiRatio} is also shown. The error bars represent the total uncertainty on the measurement, assuming isotropic decay angular distributions. }
\label{Fig:JpsiRatio}
\end{figure}

\subsection{Cross-section ratios}

The production rates of $\chicj2$ relative to $\chicj1$ are measured for prompt and non-prompt $\chic$ as a function of $\ptjpsi$. The ratio of the prompt cross-sections is shown in figure~\ref{Fig:PromptRatio}. The measurements are compared to the NLO NRQCD and CSM predictions and to the measurements of CMS within the range $|\yjpsi|<1.0$~\cite{CMS_Ratio}. The NLO NRQCD prediction is in generally good agreement with the measurements, particularly at lower $\ptjpsi$ values. The cross-section ratio predicted by the CSM is consistently lower than the measurements. The ratio of the non-prompt cross-sections is shown in figure~\ref{Fig:NonPromptRatio} and is compared to the measurement of CDF in $p\bar{p}$ collisions at $\rts=1.96 \TeV$ and for $\ptjpsi>10\GeV$~\cite{CDF_Ratio}.

\begin{figure}
\begin{center}
\includegraphics[width=0.7\textwidth]{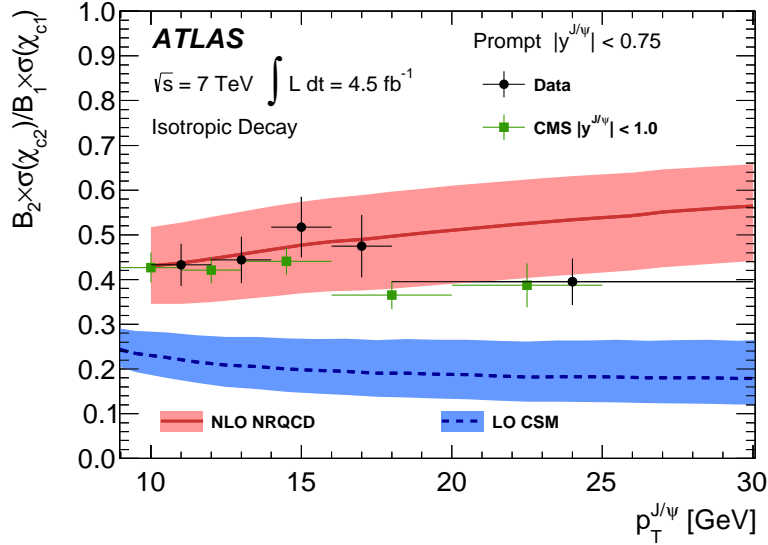}
\end{center}
\caption{The production cross-section of prompt $\chicj2$ relative to prompt $\chicj1$ measured as a function of $\ptjpsi$. The measurements are compared to the predictions of NLO NRQCD and the LO CSM. The measurement from CMS~\cite{CMS_Ratio} is also shown. The error bars represent the total uncertainty on the measurement, assuming isotropic decay angular distributions. The factors $B_{1}$ and $B_{2}$ denote the branching fractions $B_{1}={\cal B}\left(\chicj1 \to\Jpsi\,\gamma\right)$ and $B_{2}={\cal B}\left(\chicj2 \to\Jpsi\,\gamma\right)$, respectively.}
\label{Fig:PromptRatio}
\end{figure}

\begin{figure}
\begin{center}
\includegraphics[width=0.7\textwidth]{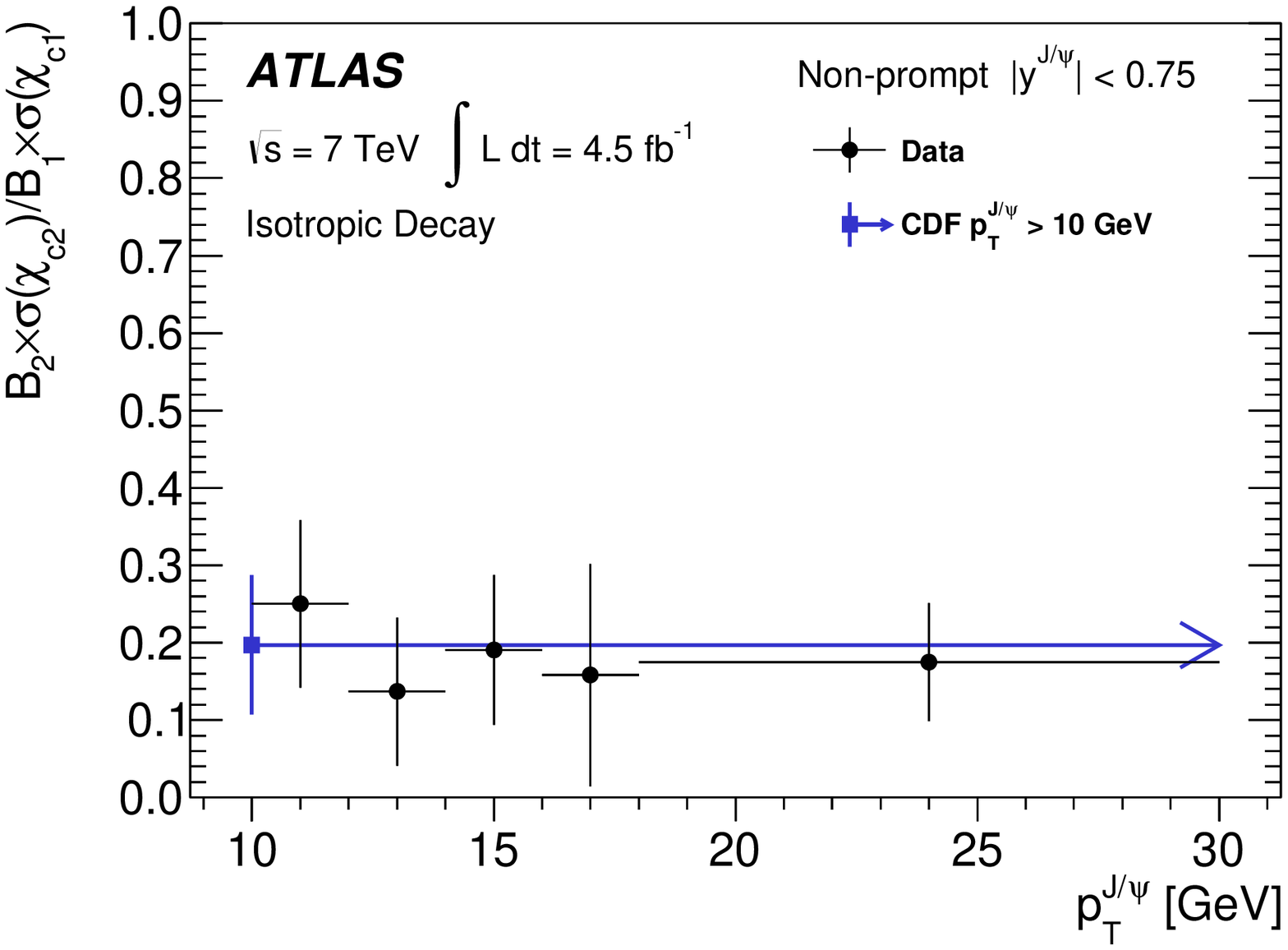}
\end{center}
\caption{The production cross-section of non-prompt $\chicj2$ relative to non-prompt $\chicj1$, ${\cal B}\left(\chicj2 \to\Jpsi\,\gamma\right)\sigma\left(\chicj2\right)/{\cal B}\left(\chicj1 \to\Jpsi\,\gamma\right)\sigma\left(\chicj1\right)$, measured as a function of $\ptjpsi$. The measurement from CDF~\cite{CDF_Ratio} is also shown. The error bars represent the total uncertainty on the measurement, assuming isotropic decay angular distributions. The factors $B_{1}$ and $B_{2}$ denote the branching fractions $B_{1}={\cal B}\left(\chicj1 \to\Jpsi\,\gamma\right)$ and $B_{2}={\cal B}\left(\chicj2 \to\Jpsi\,\gamma\right)$, respectively.}
\label{Fig:NonPromptRatio}
\end{figure}

\subsection{Non-prompt fractions}

The prompt and non-prompt $\chicj1$ and $\chicj2$ cross-sections are used to calculate the fractions of inclusive $\chicj1$ and $\chicj2$ produced in the decays of $b$-hadrons, $f_{\mathrm{non\mhyphen prompt}}$. The non-prompt fraction is measured as a function of $\ptchic$ and is shown in figure~\ref{Fig:NonPromptFraction}. The combined non-prompt fraction is observed to increase as a function of $\ptchic$, as is observed in the $\Jpsi$ and $\psi(2S)$ systems. However, the inclusive production of $\chicj1$ and $\chicj2$ is dominated by prompt production in the kinematic region measured, contrary to what is observed in the $\Jpsi$ and $\psi(2S)$ systems for the same high-$\pt$ region, where the inclusive cross-sections contain a larger component of feed-down from $b$-hadron decays~\cite{ATLAS_Jpsi,CMS_Psi2S}.

\begin{figure}
\begin{center}
\includegraphics[width=0.7\textwidth]{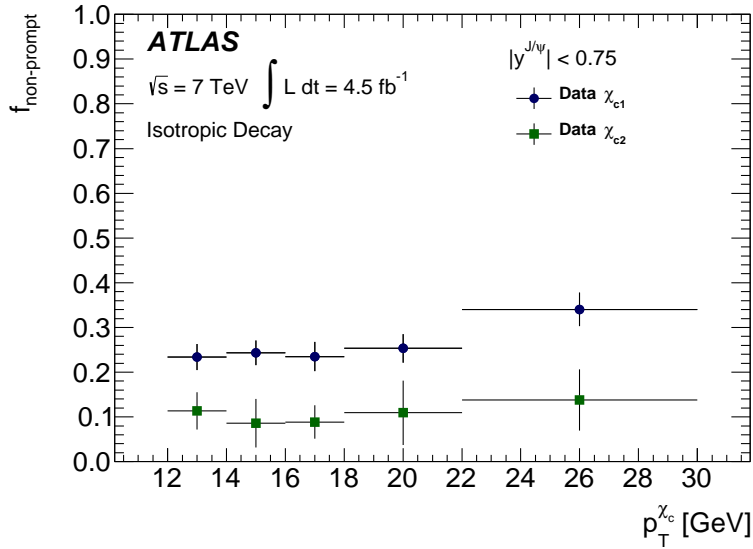}
\end{center}
\caption{The fractions of $\chicj1$ and $\chicj2$ produced in the decays of $b$-hadrons, $f_{\mathrm{non\mhyphen prompt}}$, as a function of $\ptchic$. The error bars represent the total uncertainty on the measurement, assuming isotropic decay angular distributions.}
\label{Fig:NonPromptFraction}
\end{figure}

\clearpage


\section{\boldmath Measurement of ${\cal B}\left(B^{\pm}\to\chicj1 K^{\pm}\right)$}

The branching fraction ${\cal B}\left(B^{\pm}\to\chicj1 K^{\pm}\right)$ is measured using the decay $B^{\pm}\to\Jpsi K^{\pm}$ as a reference channel (with $\chicj1\to\Jpsi\,\gamma$ and $\Jmumu$ for both channels). The final states of both channels are identical apart from the photon.

The branching fraction ${\cal B}\left(B^{\pm}\to\chicj1 K^{\pm}\right)$ is measured from
\begin{equation}
{\cal B}\left(B^{\pm}\to\chicj1 K^{\pm}\right) = {\cal A}_{B}\cdot\frac{N^{B}_{\chicj1}}{N^{B}_{\Jpsi}}\cdot\frac{{\cal B}\left(B^{\pm}\to\Jpsi K^{\pm}\right)}{{\cal B}\left(\chicj1\to\Jpsi\,\gamma\right)}\,,\nonumber
\label{Eqn:Branch}
\end{equation}

\noindent where ${\cal A}_{B}$ is a factor to correct for the different detector acceptances of the two decays, and $N^{B}_{\chicj1}$ and $N^{B}_{\Jpsi}$ are the corrected yields for the signal and reference decay channels respectively. The current world-average values are used for the branching fractions: ${\cal B}\left(B^{\pm}\to\Jpsi K^{\pm}\right) = \left(1.016\pm0.033\right)\times 10^{-3}$ and ${\cal B}\left(\chicj1\to\Jpsi\,\gamma\right) = 0.344 \pm 0.015$~\cite{PDG}. Both decays are reconstructed within the region $10 \leq \ptjpsi < 30\GeV$ and $|\yjpsi| < 0.75$.

The branching fraction ${\cal B}\left(B^{\pm}\to\chicj1 K^{\pm}\right)$ is measured using the same data sample and efficiency corrections used in the inclusive $\chic$ production measurements. The $\chicj1$ and $\Jpsi$ candidate selections (with the photon being reconstructed from conversions) and fiducial region are kept as close to the inclusive $\chic$ measurement as possible.

\subsection{\boldmath Selection of $B^{\pm}$ decays}
The selection of candidate $B^{\pm}\to\chicj1 K^{\pm}$ and $B^{\pm}\to\Jpsi K^{\pm}$ decays begins by searching for $\chic$ and $\Jpsi$ candidates using the selection criteria described in the inclusive $\chic$ measurement. Charged particles with tracks consistent with originating from the $\mumu$ vertex are assigned the charged kaon mass and the $\mumu K^{\pm}$ vertex is fitted. The candidate charged kaon track is required to contain at least one silicon pixel hit and at least six SCT hits, transverse momentum $\pt>3\GeV$ and pseudorapidity $|\eta|<2.5$. Candidates with a vertex fit quality $\chi^{2}$ per degree of freedom $< 6$ are retained. The $L_{xy}$ of the $\Jmumu$ candidate (as defined in section~\ref{Sec:CrossSection}) is required to be greater than $0.3\,\mathrm{mm}$ to reject promptly produced $\Jpsi$ mesons. This cut rejects over $99\%$ of the prompt $\Jpsi$ background and retains around $87\%$ of the $B^{\pm}$ signal. Candidate $B^{\pm}\to\chicj1 K^{\pm}\to\mumu\gamma K^{\pm}$ decays are required to have $0.32 < m\left(\mumu\gamma\right) - m\left(\mumu\right) < 0.43\GeV$ to select $\chicj1$ decays and $4.65 < m\left(\mumu K^{\pm}\right) - m\left(\mumu\right) + m_{\Jpsi} < 5.2\GeV$ to reject backgrounds from $B^{\pm}\to\Jpsi K^{\pm}$ decays.

\subsection{\boldmath Calculation of ${\cal A}_{B}$}

The acceptance correction factor ${\cal A}_{B}$ is defined as the number of $B^{\pm}\to\Jpsi K^{\pm}$ decays relative to the number of $B^{\pm}\to\chicj1 K^{\pm}$ decays that fall within their respective fiducial regions. The correction is derived from a large sample of generator-level MC simulation events that uses a fitted parameterisation of the ATLAS measurement of the $B^{\pm}$ differential cross-section~\cite{ATLAS_BPlus} (in the $B^{\pm}\to\Jpsi K^{\pm}$ mode) as an input. The simulation generates $B^{\pm}\to\Jpsi K^{\pm}$ and $B^{\pm}\to\chicj1 K^{\pm}$ decays according to the measured spectrum. The angular distributions of the $B^{\pm}$ decay products are generated with helicity equal to zero for the charmonium state in the rest frame of the $B^{\pm}$ meson. This calculation gives ${\cal A}_{B} = 2.30\pm0.08$ where the uncertainty is derived from the uncertainty in the fitted parameterisation of the measured $B^{\pm}$ cross-section. The difference in the values of ${\cal A}_{B}$ calculated with the nominal and alternative fit parameterisations is taken as an estimate of the systematic uncertainty.

\subsection{\boldmath Extraction of $N^{B}_{\chicj1}$ and $N^{B}_{\Jpsi}$}

Candidate $B^{\pm}\to\chicj1 K^{\pm}$ and $B^{\pm}\to\Jpsi K^{\pm}$ decays are weighted to correct for trigger efficiency, muon reconstruction efficiency and (for $B^{\pm}\to\chicj1 K^{\pm}$ decays) conversion probability and converted-photon reconstruction efficiency using the same corrections derived for the inclusive $\chic$ production measurement. Corrections are applied only for effects that are known not to fully cancel in the ratio $N^{B}_{\chicj1}/N^{B}_{\Jpsi}$. Trigger and muon reconstruction efficiencies are corrected since not all $\Jmumu$ decays in the fiducial region studied fall within the efficiency plateau. Since the data are only partially corrected, the weighted yields are not representative of the true yields of $B^{\pm}$ mesons one would expect from the data sample and fiducial region used. The corrected yields $N^{B}_{\chicj1}$ and $N^{B}_{\Jpsi}$ are extracted from unbinned maximum likelihood fits to the $m\left(\mumu\gamma K^{\pm}\right) - m\left(\mumu\gamma\right) + m_{\chicj1}$ and $m\left(\mumu K^{\pm}\right) - m\left(\mumu\right) + m_{\Jpsi}$ distributions of selected $B^{\pm}$ decay candidates, where $m_{\Jpsi}$ and $m_{\chicj1}$ are the world-average values for the masses of the $\Jpsi$ and $\chicj1$ states~\cite{PDG}.

The mass distribution for candidate $B^{\pm}\to\chicj1 K^{\pm}$ decays is fitted with a $B^{\pm}$ signal modelled by a Gaussian pdf where both the mean value and width are free parameters in the fit. The background distribution is modelled with a template derived from MC simulation of inclusive $pp\to\bbbar X$ decays generated by {\sc Pythia 6}~\cite{Pythia6} and processed with the detector simulation. The Gaussian kernel estimation~\cite{GaussKernel} procedure is applied to the background template from MC simulation to form a non-analytic background pdf. The mass distribution for candidate $B^{\pm}\to\Jpsi K^{\pm}$ decays is modelled with a double-Gaussian signal pdf where the mean value (common to both Gaussian pdfs), both width parameters and the relative normalisation of both components are determined by the fit. The background contribution to the $B^{\pm}\to\Jpsi K^{\pm}$ mass distribution from $B^{\pm}\to\chicj{1,2}K^{\pm}$ and $B^{\pm/0}\to\Jpsi\left(K\pi\right)^{\pm/0}$ decays (where only the $\Jpsi$ and charged kaon are reconstructed) is modelled by the sum of a Gaussian and a complementary error function ~\cite{ATLAS_BPlus}. The background contribution from $B^{\pm}\to\Jpsi\pi^{\pm}$ decays where the kaon mass is wrongly assigned to the pion track is modelled with a CB function~\cite{ATLAS_BPlus}.

The results of the fits to the mass distributions of candidate $B^{\pm}\to\chicj1 K^{\pm}$ and $B^{\pm}\to\Jpsi K^{\pm}$ decays are shown in figure~\ref{Fig:BFits}.

\begin{figure}
\begin{center}
\includegraphics[width=0.7\textwidth]{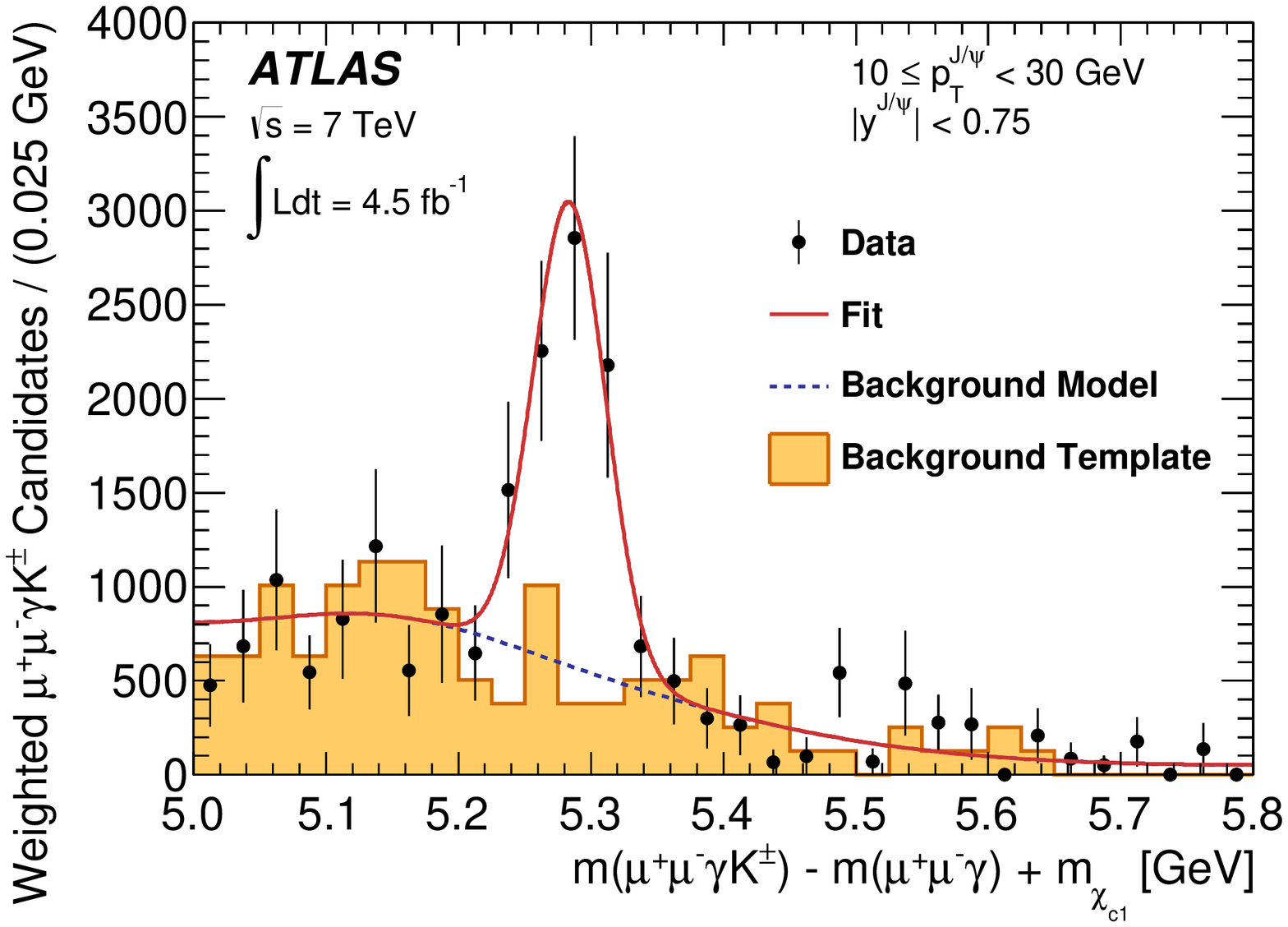}

\includegraphics[width=0.7\textwidth]{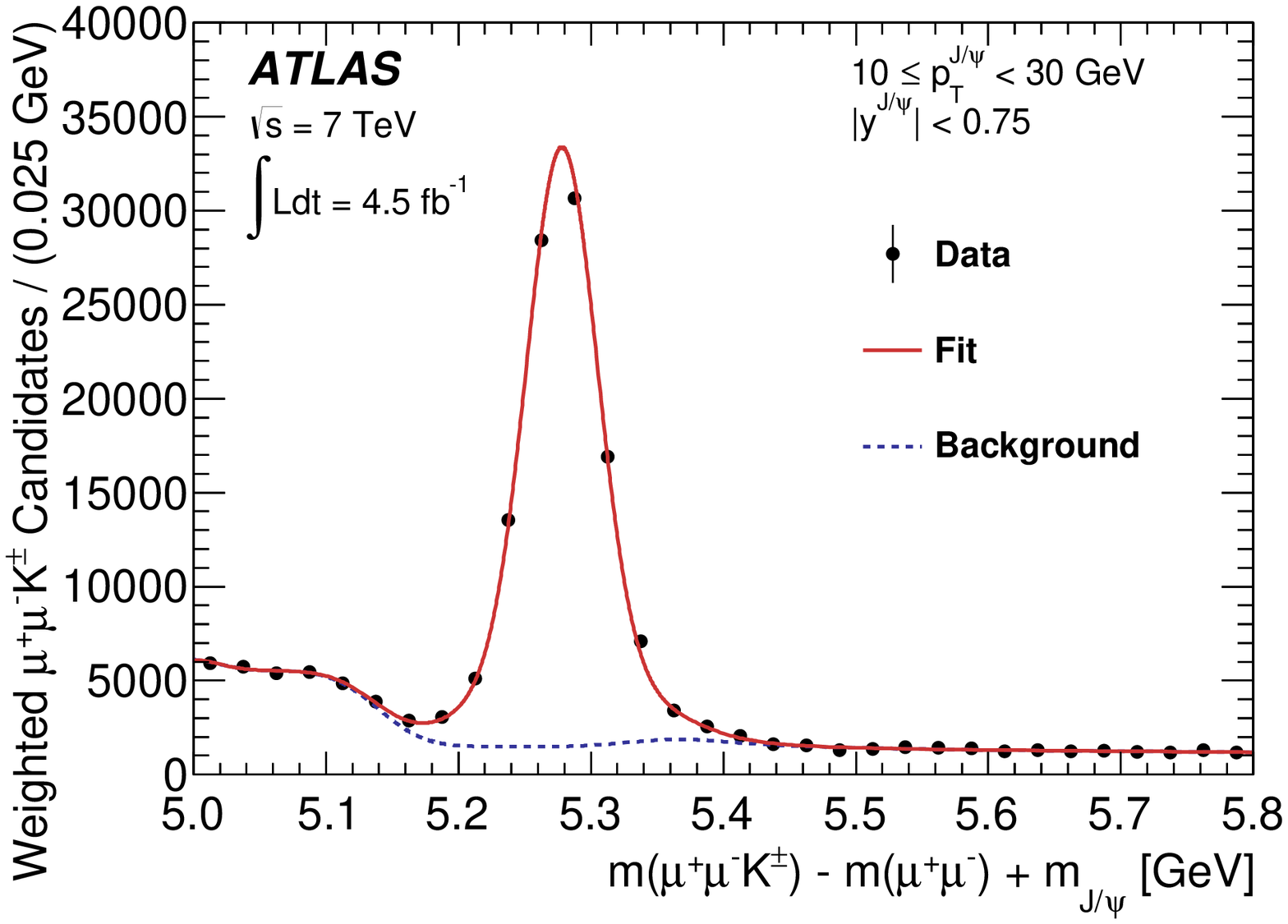}
\end{center}
\caption{The results of fits to the $m\left(\mumu\gamma K^{\pm}\right) - m\left(\mumu\gamma\right) + m_{\chicj1}$ (top) and $m\left(\mumu K^{\pm}\right) - m\left(\mumu\right) + m_{\Jpsi}$ (bottom) distributions of selected $B^{\pm}$ decay candidates. The background template derived from MC simulation is shown as the shaded histogram in the top figure. }
\label{Fig:BFits}
\end{figure}

\subsection{Systematic uncertainties}
Several sources of systematic uncertainty on the measurement are considered. The systematic uncertainties due to the efficiency corrections (trigger, conversion probability, and muon and conversion reconstruction) are quantified with the same methods used in the inclusive $\chic$ measurement. Several variations in both fit models are also tested to estimate the systematic uncertainty on $N^{B}_{\chicj1}/N^{B}_{\Jpsi}$ due to the fit models. The systematic uncertainty is taken as the maximum deviation of any single combination of alternative fit results from the average. The systematic uncertainty on ${\cal A}_{B}$ due to the fitted parameterisation of the $B^{\pm}$ cross-section is also propagated into an uncertainty on ${\cal B}\left(B^{\pm}\to\chicj1 K^{\pm}\right)$. Table~\ref{Tab:Branch_Syst} shows a summary of the individual sources of systematic uncertainty considered.

\begin{table}[h]
\begin{center}
\begin{tabular}{| l | c |}
\hline
& Fractional Uncertainty [$\%$]\\
\hline                        
Converted-photon reconstruction efficiency & 10 \\
Conversion probability & 4 \\
Muon reconstruction efficiency & 1 \\
Trigger efficiency & 1 \\
Acceptance & 3 \\
Fit model & 6 \\
\hline
Statistical & 18 \\
Systematic & 13 \\
\hline
Total & 22 \\
\hline
\end{tabular}
\end{center}

\caption{Sources of systematic uncertainty on the measurement of ${\cal B}(B^{\pm}\to\chicj1 K^{\pm})$.}
\label{Tab:Branch_Syst}
\end{table}

\subsection{Result}

The measured branching fraction is ${\cal B}(B^{\pm}\to\chicj1 K^{\pm}) = (4.9 \pm 0.9 \stat \pm 0.6 \syst)\times 10^{-4}$. This value is in good agreement with the current world-average value of $(4.79\pm0.23)\times 10^{-4}$~\cite{PDG} (dominated by measurements from Belle~\cite{BelleBranch} and BaBar~\cite{BabarBranch}), and supports the estimate of the conversion reconstruction efficiencies at the level of $19\%$ (the total fractional uncertainty on the measurement, neglecting conversion-related systematic uncertainties). The precision of this measurement is significantly better than previous measurements from hadron collider experiments~\cite{PDG}.


\section{Conclusion}

The cross-sections for prompt and non-prompt $\chicj1$ and $\chicj2$ production have been measured in $4.5\,\ifb$ of $pp$ collisions at $\rts=7\TeV$ with the ATLAS detector at the LHC. The $\chic$ states are reconstructed from the radiative decay $\chic\to\Jpsi\,\gamma$. The measurements are performed as a function of both $\ptjpsi$ and $\ptchic$ within the rapidity interval $|\yjpsi| < 0.75$. The results, measured as a function of $\ptjpsi$ and $\ptchic$, are presented within the regions $10\leq\ptjpsi<30\GeV$ and $12\leq\ptchic<30\GeV$, respectively. The production rate of the $\chicj2$ state is measured relative to the $\chicj1$ state for both prompt and non-prompt production as a function of $\ptjpsi$. The measurements of prompt $\chic$ are combined with existing ATLAS measurements of prompt $\Jpsi$ production to derive the fraction of prompt $\Jpsi$ produced in feed-down from $\chic$ decays. The fractions of $\chicj1$ and $\chicj2$ produced in the decays of $b$-hadrons are also presented as functions of $\ptchic$.

The measurements of prompt $\chic$ production are compared to the theoretical predictions of NLO NRQCD, the $k_{\mathrm{T}}$ factorisation approach and the Colour Singlet Model. The NRQCD predictions generally agree well with the data. The $k_{\mathrm{T}}$ factorisation approach predicts a cross-section significantly in excess of the measurement while the CSM prediction significantly underestimates the data. This suggests that higher-order corrections or colour-octet contributions to the cross-sections not included in either prediction may be numerically important. The measurements of non-prompt $\chic$ production generally agree well with predictions based upon the FONLL approach.

The branching fraction ${\cal B}(B^{\pm}\to\chicj1 K^{\pm}) = (4.9 \pm 0.9 \stat \pm 0.6 \syst)\times 10^{-4}$ is also measured with the same dataset and $\chic$ event selection. The measured value agrees well with the world average and supports the estimate of the conversion reconstruction efficiencies derived from simulation.


\section*{Acknowledgements}

We thank CERN for the very successful operation of the LHC, as well as the
support staff from our institutions without whom ATLAS could not be
operated efficiently.

We acknowledge the support of ANPCyT, Argentina; YerPhI, Armenia; ARC,
Australia; BMWF and FWF, Austria; ANAS, Azerbaijan; SSTC, Belarus; CNPq and FAPESP,
Brazil; NSERC, NRC and CFI, Canada; CERN; CONICYT, Chile; CAS, MOST and NSFC,
China; COLCIENCIAS, Colombia; MSMT CR, MPO CR and VSC CR, Czech Republic;
DNRF, DNSRC and Lundbeck Foundation, Denmark; EPLANET, ERC and NSRF, European Union;
IN2P3-CNRS, CEA-DSM/IRFU, France; GNSF, Georgia; BMBF, DFG, HGF, MPG and AvH
Foundation, Germany; GSRT and NSRF, Greece; ISF, MINERVA, GIF, I-CORE and Benoziyo Center,
Israel; INFN, Italy; MEXT and JSPS, Japan; CNRST, Morocco; FOM and NWO,
Netherlands; BRF and RCN, Norway; MNiSW and NCN, Poland; GRICES and FCT, Portugal; MNE/IFA, Romania; MES of Russia and ROSATOM, Russian Federation; JINR; MSTD,
Serbia; MSSR, Slovakia; ARRS and MIZ\v{S}, Slovenia; DST/NRF, South Africa;
MINECO, Spain; SRC and Wallenberg Foundation, Sweden; SER, SNSF and Cantons of
Bern and Geneva, Switzerland; NSC, Taiwan; TAEK, Turkey; STFC, the Royal
Society and Leverhulme Trust, United Kingdom; DOE and NSF, United States of
America.

The crucial computing support from all WLCG partners is acknowledged
gratefully, in particular from CERN and the ATLAS Tier-1 facilities at
TRIUMF (Canada), NDGF (Denmark, Norway, Sweden), CC-IN2P3 (France),
KIT/GridKA (Germany), INFN-CNAF (Italy), NL-T1 (Netherlands), PIC (Spain),
ASGC (Taiwan), RAL (UK) and BNL (USA) and in the Tier-2 facilities
worldwide.

\providecommand{\href}[2]{#2}\begingroup\raggedright
\endgroup

\clearpage

\appendix
\section{Tabulated results}
\label{Sec:Appendix}

The following tables show the results presented in section\,\ref{Sec:Results}. Statistical and systematic uncertainties are shown for each measurement along with the uncertainty envelope associated with the unknown $\chic$ spin alignment.

\begin{table}[h]
\begin{center}
\begin{tabular}{| c | c | c | c | c | c | c |}
\hline\hline
\multicolumn{2}{|c|}{} & \multicolumn{5}{|c|}{\multirow{2}{*}{${\cal B}\left(\chicj J\to\Jpsi\,\gamma\right)\cdot{\cal B}\left(\Jmumu\right)\cdot \frac{\mathrm{d}\sigma_{J}^{\mathrm{P}}}{\mathrm{d}\pt}\,[\mathrm{pb/GeV}]$}} \\
\multicolumn{2}{|c|}{} & \multicolumn{5}{|c|}{} \\
\hline
$\ptjpsi\,[\mathrm{GeV}]$ & $\langle\ptjpsi\rangle\,[\mathrm{GeV}]$ & $J$ & Value & $\stat$ & $\syst$ & Spin-alignment envelope \\
\hline
\multirow{2}{*}{$10.0$--$12.0$} & $11.0$ & $1$ & $218$ &$\pm9$ & $\pm28$ & $+69$ $-28$ \\ 
& $11.0$ & $2$ & $95$ &$\pm6$ & $\pm12$ & $+34$ $-21$ \\ 
\hline
\multirow{2}{*}{$12.0$--$14.0$} & $12.9$ & $1$ & $90$ &$\pm4$ & $\pm11$ & $+31$ $-12$ \\ 
& $12.9$ & $2$ & $40$ &$\pm3$ & $\pm5$ & $+15$ $-10$ \\ 
\hline
\multirow{2}{*}{$14.0$--$16.0$} & $14.9$ & $1$ & $37$ &$\pm2$ & $\pm5$ & $+13$ $-5$ \\ 
& $14.9$ & $2$ & $19$ &$\pm2$ & $\pm2$ & $+7$ $-5$ \\ 
\hline
\multirow{2}{*}{$16.0$--$18.0$} & $16.9$ & $1$ & $21$ &$\pm1$ & $\pm3$ & $+7$ $-3$ \\ 
& $16.9$ & $2$ & $10$ &$\pm1$ & $\pm1$ & $+4$ $-2$ \\ 
\hline
\multirow{2}{*}{$18.0$--$30.0$} & $22.1$ & $1$ & $4.8$ &$\pm0.2$ & $\pm0.6$ & $+1.5$ $-0.6$ \\ 
& $22.1$ & $2$ & $1.9$ &$\pm0.2$ & $\pm0.2$ & $+0.6$ $-0.4$ \\ 
\hline
\hline
\end{tabular}
\end{center}
\caption{Differential cross-section for prompt $\chicj1$ and $\chicj2$ production, measured in bins of $\ptjpsi$.}
\label{Tab:PromptCS_JpsiPt}
\end{table}

\begin{table}[h]
\begin{center}
\begin{tabular}{| c | c | c | c | c | c | c |}
\hline\hline
\multicolumn{2}{|c|}{} & \multicolumn{5}{|c|}{\multirow{2}{*}{${\cal B}\left(\chicj J\to\Jpsi\,\gamma\right)\cdot{\cal B}\left(\Jmumu\right)\cdot \frac{\mathrm{d}\sigma_{J}^{\mathrm{NP}}}{\mathrm{d}\pt}\,[\mathrm{pb/GeV}]$}} \\
\multicolumn{2}{|c|}{} & \multicolumn{5}{|c|}{} \\
\hline
$\ptjpsi\,[\mathrm{GeV}]$ & $\langle\ptjpsi\rangle\,[\mathrm{GeV}]$ & $J$ & Value & $\stat$ & $\syst$ & Spin-alignment envelope \\
\hline
\multirow{2}{*}{$10.0$--$12.0$}& $11.0$ & $1$ &  $60$ &$\pm8$ & $\pm8$ & $+19$ $-8$ \\ 
 & $11.0$& $2$ &  $15$ &$\pm6$ & $\pm3$ & $+5$ $-3$ \\ 
\hline
\multirow{2}{*}{$12.0$--$14.0$}& $12.9$ & $1$ &  $30$ &$\pm3$ & $\pm4$ & $+10$ $-4$ \\ 
 & $12.9$& $2$ &  $4.1$ &$\pm2.8$ & $\pm0.7$ & $+1.6$ $-1.0$ \\ 
\hline
\multirow{2}{*}{$14.0$--$16.0$}& $14.9$ & $1$ &  $15$ &$\pm2$ & $\pm2$ & $+5$ $-2$ \\ 
 & $14.9$& $2$ &  $2.9$ &$\pm1.3$ & $\pm0.5$ & $+1.1$ $-0.7$ \\ 
\hline
\multirow{2}{*}{$16.0$--$18.0$}& $16.9$ & $1$ &  $5.8$ &$\pm1.1$ & $\pm0.8$ & $+1.9$ $-0.7$ \\ 
 & $16.9$& $2$ &  $0.9$ &$\pm0.8$ & $\pm0.2$ & $+0.3$ $-0.2$ \\ 
\hline
\multirow{2}{*}{$18.0$--$30.0$}& $22.1$ & $1$ &  $2.1$ &$\pm0.3$ & $\pm0.3$ & $+0.6$ $-0.2$ \\ 
 & $22.1$& $2$ &  $0.4$ &$\pm0.1$ & $\pm0.1$ & $+0.1$ $-0.1$ \\ 
\hline
\hline
\end{tabular}
\end{center}
\caption{Differential cross-section for non-prompt $\chicj1$ and $\chicj2$ production, measured in bins of $\ptjpsi$.}
\label{Tab:NonPromptCS_JpsiPt}
\end{table}

\begin{table}[h]
\begin{center}
\begin{tabular}{| c | c | c | c | c | c | c |}
\hline\hline
\multicolumn{2}{|c|}{} & \multicolumn{5}{|c|}{\multirow{2}{*}{${\cal B}\left(\chicj J\to\Jpsi\,\gamma\right)\cdot{\cal B}\left(\Jmumu\right)\cdot \frac{\mathrm{d}\sigma_{J}^{\mathrm{P}}}{\mathrm{d}\pt}\,[\mathrm{pb/GeV}]$}} \\
\multicolumn{2}{|c|}{} & \multicolumn{5}{|c|}{}\\
\hline
$\ptchic\,[\mathrm{GeV}]$ & $\langle\ptchic\rangle\,[\mathrm{GeV}]$ & $J$ & Value & $\stat$ & $\syst$ & Spin-alignment envelope \\
\hline
\multirow{2}{*}{$12.0$--$14.0$} & $12.9$ & $1$ & $136$ &$\pm7$ & $\pm16$ & $+41$ $-17$ \\ 
& $12.9$ & $2$ & $73$ &$\pm5$ & $\pm9$ & $+25$ $-15$ \\ 
\hline
\multirow{2}{*}{$14.0$--$16.0$} & $14.9$ & $1$ & $71$ &$\pm3$ & $\pm9$ & $+22$ $-8$ \\ 
& $14.9$ & $2$ & $29$ &$\pm2$ & $\pm4$ & $+10$ $-6$ \\ 
\hline
\multirow{2}{*}{$16.0$--$18.0$} & $16.9$ & $1$ & $31$ &$\pm2$ & $\pm4$ & $+10$ $-4$ \\ 
& $16.9$ & $2$ & $18$ &$\pm1$ & $\pm2$ & $+6$ $-4$ \\ 
\hline
\multirow{2}{*}{$18.0$--$22.0$} & $19.6$ & $1$ & $15.4$ &$\pm0.8$ & $\pm1.8$ & $+4.4$ $-1.8$ \\ 
& $19.7$ & $2$ & $7.0$ &$\pm0.6$ & $\pm0.9$ & $+2.1$ $-1.4$ \\ 
\hline
\multirow{2}{*}{$22.0$--$30.0$} & $25.0$ & $1$ & $4.0$ &$\pm0.2$ & $\pm0.5$ & $+1.0$ $-0.4$ \\ 
& $25.0$ & $2$ & $1.7$ &$\pm0.2$ & $\pm0.2$ & $+0.4$ $-0.3$ \\ 
\hline
\hline
\end{tabular}
\end{center}
\caption{Differential cross-section for prompt $\chicj1$ and $\chicj2$ production, measured in bins of $\ptchic$.}
\label{Tab:PromptCS_ChiPt}
\end{table}

\begin{table}[h]
\begin{center}
\begin{tabular}{| c | c | c | c | c | c | c |}
\hline\hline
\multicolumn{2}{|c|}{} & \multicolumn{5}{|c|}{\multirow{2}{*}{${\cal B}\left(\chicj J\to\Jpsi\,\gamma\right)\cdot{\cal B}\left(\Jmumu\right)\cdot \frac{\mathrm{d}\sigma_{J}^{\mathrm{NP}}}{\mathrm{d}\pt}\,[\mathrm{pb/GeV}]$}} \\
\multicolumn{2}{|c|}{} & \multicolumn{5}{|c|}{} \\
\hline
$\ptchic\,[\mathrm{GeV}]$ & $\langle\ptchic\rangle\,[\mathrm{GeV}]$ & $J$ & Value & $\stat$ & $\syst$ & Spin-alignment envelope \\
\hline
\multirow{2}{*}{$12.0$--$14.0$}& $12.9$ & $1$ &  $42$ &$\pm6$ & $\pm5$ & $+13$ $-5$ \\ 
 & $12.9$& $2$ &  $9$ &$\pm4$ & $\pm1$ & $+3$ $-2$ \\ 
\hline
\multirow{2}{*}{$14.0$--$16.0$}& $14.9$ & $1$ &  $23$ &$\pm2$ & $\pm3$ & $+7$ $-3$ \\ 
 & $14.9$& $2$ &  $2.7$ &$\pm1.8$ & $\pm0.4$ & $+0.9$ $-0.6$ \\ 
\hline
\multirow{2}{*}{$16.0$--$18.0$}& $16.9$ & $1$ &  $10$ &$\pm2$ & $\pm1$ & $+3$ $-1$ \\ 
 & $16.9$& $2$ &  $1.8$ &$\pm0.9$ & $\pm0.3$ & $+0.6$ $-0.4$ \\ 
\hline
\multirow{2}{*}{$18.0$--$22.0$}& $19.6$ & $1$ &  $5.2$ &$\pm0.8$ & $\pm0.6$ & $+1.5$ $-0.6$ \\ 
 & $19.7$& $2$ &  $0.9$ &$\pm0.5$ & $\pm0.1$ & $+0.3$ $-0.2$ \\ 
\hline
\multirow{2}{*}{$22.0$--$30.0$}& $25.0$ & $1$ &  $2.1$ &$\pm0.3$ & $\pm0.3$ & $+0.5$ $-0.2$ \\ 
 & $25.0$& $2$ &  $0.27$ &$\pm0.17$ & $\pm0.04$ & $+0.07$ $-0.05$ \\ 
\hline\hline
\end{tabular}
\end{center}
\caption{Differential cross-section for non-prompt $\chicj1$ and $\chicj2$ production, measured in bins of $\ptchic$.}
\label{Tab:NonPromptCS_ChiPt}
\end{table}

\begin{table}[h]
\begin{center}
\begin{tabular}{| c | c | c | c | c |}
\hline\hline
& \multicolumn{4}{|c|}{\multirow{2}{*}{Prompt $R_{\chic}$}} \\
& \multicolumn{4}{|c|}{} \\
\hline
$\ptjpsi\,[\mathrm{GeV}]$ & Value & $\stat$ & $\syst$ & Spin-alignment envelope \\
\hline
$10.0$--$12.0$ & $0.24$ &$\pm0.02$ & $\pm0.04$ & $+0.08$ $-0.04$ \\ 
\hline
$12.0$--$14.0$ & $0.26$ &$\pm0.01$ & $\pm0.04$ & $+0.09$ $-0.04$ \\ 
\hline
$14.0$--$16.0$ & $0.23$ &$\pm0.02$ & $\pm0.04$ & $+0.08$ $-0.04$ \\ 
\hline
$16.0$--$18.0$ & $0.28$ &$\pm0.03$ & $\pm0.05$ & $+0.10$ $-0.05$ \\ 
\hline
$18.0$--$30.0$ & $0.27$ &$\pm0.03$ & $\pm0.05$ & $+0.09$ $-0.04$ \\ 
\hline
\hline
\end{tabular}
\end{center}
\caption{Fraction of prompt $\Jpsi$ produced in feed-down from $\chic$ decays as a function of $\ptjpsi$. The spin alignment envelope assumes that prompt $\Jpsi$ are produced unpolarised and represents the maximum uncertainty in the result due to the unknown $\chic$ spin alignment.}
\label{Tab:JpsiRatio}
\end{table}

\begin{table}[h]
\begin{center}
\begin{tabular}{| c | c | c | c | c |}
\hline\hline
& \multicolumn{4}{|c|}{\multirow{2}{*}{Prompt $\frac{\sigma\left(\chicj2\right)\cdot{\cal B}\left(\chicj2\to\Jpsi\,\gamma\right)}{\sigma\left(\chicj1\right)\cdot{\cal B}\left(\chicj1\to\Jpsi\,\gamma\right)}$}} \\
& \multicolumn{4}{|c|}{}\\
\hline
$\ptjpsi\,[\mathrm{GeV}]$ & Value & $\stat$ & $\syst$ & Spin-alignment envelope \\
\hline
$10.0$--$12.0$ & $0.43$ &$\pm0.04$ & $\pm0.03$ & $+0.24$ $-0.18$ \\ 
\hline
$12.0$--$14.0$ & $0.44$ &$\pm0.04$ & $\pm0.03$ & $+0.26$ $-0.19$ \\ 
\hline
$14.0$--$16.0$ & $0.52$ &$\pm0.06$ & $\pm0.04$ & $+0.30$ $-0.23$ \\ 
\hline
$16.0$--$18.0$ & $0.48$ &$\pm0.06$ & $\pm0.03$ & $+0.27$ $-0.21$ \\ 
\hline
$18.0$--$30.0$ & $0.40$ &$\pm0.04$ & $\pm0.03$ & $+0.20$ $-0.16$ \\ 
\hline\hline
\end{tabular}
\end{center}
\caption{Production rate of prompt $\chicj2$ relative to prompt $\chicj1$, measured in bins of $\ptjpsi$.}
\label{Tab:PromptRatio}
\end{table}

\begin{table}[h]
\begin{center}
\begin{tabular}{| c | c | c | c | c |}
\hline\hline
& \multicolumn{4}{|c|}{\multirow{2}{*}{Non-prompt $\frac{\sigma\left(\chicj2\right)\cdot{\cal B}\left(\chicj2\to\Jpsi\,\gamma\right)}{\sigma\left(\chicj1\right)\cdot{\cal B}\left(\chicj1\to\Jpsi\,\gamma\right)}$}} \\
& \multicolumn{4}{|c|}{}\\
\hline
$\ptjpsi\,[\mathrm{GeV}]$ & Value & $\stat$ & $\syst$ & Spin-alignment envelope \\
\hline
$10.0$--$12.0$ & $0.25$ &$\pm0.10$ & $\pm0.03$ & $+0.14$ $-0.10$ \\ 
\hline
$12.0$--$14.0$ & $0.14$ &$\pm0.09$ & $\pm0.02$ & $+0.08$ $-0.06$ \\ 
\hline
$14.0$--$16.0$ & $0.19$ &$\pm0.09$ & $\pm0.02$ & $+0.11$ $-0.08$ \\ 
\hline
$16.0$--$18.0$ & $0.16$ &$\pm0.14$ & $\pm0.02$ & $+0.09$ $-0.07$ \\ 
\hline
$18.0$--$30.0$ & $0.18$ &$\pm0.07$ & $\pm0.02$ & $+0.09$ $-0.07$ \\ 
\hline\hline
\end{tabular}
\end{center}
\caption{Production rate of non-prompt $\chicj2$ relative to non-prompt $\chicj1$, measured in bins of $\ptjpsi$.}
\label{Tab:NonPromptRatio}
\end{table}

\begin{table}[h]
\begin{center}
\begin{tabular}{| c | c | c | c | c | c |}
\hline\hline
& \multicolumn{5}{|c|}{\multirow{2}{*}{$f_{\mathrm{non\mhyphen prompt}}$}} \\
& \multicolumn{5}{|c|}{} \\
\hline
$\ptchic\,[\mathrm{GeV}]$ & $J$ & Value & $\stat$ & $\syst$ & Spin-alignment envelope \\
\hline
\multirow{2}{*}{$12.0$--$14.0$}& $1$ & $0.23$ &$\pm0.02$ & $\pm0.02$ & $+0.08$ $-0.07$ \\ 
& $2$ & $0.11$ &$\pm0.04$ & $\pm0.01$ & $+0.07$ $-0.04$ \\ 
\hline
\multirow{2}{*}{$14.0$--$16.0$}& $1$ & $0.24$ &$\pm0.02$ & $\pm0.02$ & $+0.08$ $-0.07$ \\ 
& $2$ & $0.09$ &$\pm0.05$ & $\pm0.01$ & $+0.05$ $-0.03$ \\ 
\hline
\multirow{2}{*}{$16.0$--$18.0$}& $1$ & $0.2$ &$\pm0.03$ & $\pm0.02$ & $+0.08$ $-0.06$ \\ 
& $2$ & $0.09$ &$\pm0.04$ & $\pm0.01$ & $+0.05$ $-0.03$ \\ 
\hline
\multirow{2}{*}{$18.0$--$22.0$}& $1$ & $0.25$ &$\pm0.03$ & $\pm0.02$ & $+0.08$ $-0.07$ \\ 
& $2$ & $0.11$ &$\pm0.07$ & $\pm0.01$ & $+0.06$ $-0.04$ \\ 
\hline
\multirow{2}{*}{$22.0$--$30.0$}& $1$ & $0.34$ &$\pm0.03$ & $\pm0.03$ & $+0.08$ $-0.07$ \\ 
& $2$ & $0.14$ &$\pm0.07$ & $\pm0.02$ & $+0.06$ $-0.04$ \\ 
\hline
\hline
\end{tabular}
\end{center}
\caption{Fraction of $\chicj1$ and $\chicj2$ produced in $b$-hadron decays as a function of $\ptchic$.}
\label{Tab:NonPromptFraction}
\end{table}

\begin{table}[h]
\begin{center}
\begin{tabular}{| c | c | c | c | c | c | c |}
\hline
Bin & Yield & Helicity $0$ & Helicity $\pm1$ & Helicity $\pm2$ & $\mathrm{AZ+}$ & $\mathrm{AZ-}$ \\
\hline
\multirow{4}{*}{$10 \leq \ptjpsi < 12\GeV$} & P1  & $1.32$ & $0.89$ & $-$ & $0.91$ & $0.87$ \\
& P2  & $0.78$ & $0.88$ & $1.35$ & $1.10$ & $1.04$ \\
& NP1 & $1.31$ & $0.89$ & $-$ & $0.91$ & $0.87$ \\
& NP2 & $0.77$ & $0.87$ & $1.37$ & $1.11$ & $1.04$ \\
\hline
\multirow{4}{*}{$12 \leq \ptjpsi < 14\GeV$} & P1  & $1.34$ & $0.88$ & $-$ & $0.89$ & $0.87$ \\
& P2  & $0.76$ & $0.87$ & $1.38$ & $1.10$ & $1.06$ \\
& NP1 & $1.33$ & $0.88$ & $-$ & $0.89$ & $0.87$ \\
& NP2 & $0.76$ & $0.87$ & $1.38$ & $1.10$ & $1.06$ \\
\hline
\multirow{4}{*}{$14 \leq \ptjpsi < 16\GeV$} & P1  & $1.35$ & $0.88$ & $-$ & $0.88$ & $0.87$ \\
& P2  & $0.76$ & $0.87$ & $1.38$ & $1.09$ & $1.07$ \\
& NP1 & $1.34$ & $0.88$ & $-$ & $0.89$ & $0.87$ \\
& NP2 & $0.76$ & $0.87$ & $1.38$ & $1.09$ & $1.07$ \\
\hline
\multirow{4}{*}{$16 \leq \ptjpsi < 18\GeV$} & P1  & $1.35$ & $0.88$ & $-$ & $0.88$ & $0.87$ \\
& P2  & $0.76$ & $0.87$ & $1.37$ & $1.09$ & $1.07$ \\
& NP1 & $1.33$ & $0.88$ & $-$ & $0.88$ & $0.87$ \\
& NP2 & $0.76$ & $0.87$ & $1.37$ & $1.09$ & $1.07$ \\
\hline
\multirow{4}{*}{$18 \leq \ptjpsi < 30\GeV$} & P1  & $1.32$ & $0.88$ & $-$ & $0.89$ & $0.88$ \\
& P2  & $0.78$ & $0.88$ & $1.33$ & $1.08$ & $1.07$ \\
& NP1 & $1.30$ & $0.89$ & $-$ & $0.89$ & $0.88$ \\
& NP2 & $0.78$ & $0.88$ & $1.33$ & $1.07$ & $1.06$ \\
\hline
\hline
\multirow{4}{*}{$12 \leq \ptchic < 14\GeV$} & P1 & $1.31$ & $0.89$ & $-$ & $0.91$ & $0.87$ \\
& P2 & $0.78$ & $0.88$ & $1.35$ & $1.10$ & $1.04$ \\
& NP1 & $1.31$ & $0.89$ & $-$ & $0.91$ & $0.87$ \\
& NP2 & $0.78$ & $0.88$ & $1.35$ & $1.10$ & $1.04$ \\
\hline
\multirow{4}{*}{$14 \leq \ptchic < 16\GeV$} & P1 & $1.32$ & $0.89$ & $-$ & $0.90$ & $0.88$ \\
& P2 & $0.78$ & $0.88$ & $1.35$ & $1.09$ & $1.05$ \\
& NP1 & $1.32$ & $0.89$ & $-$ & $0.90$ & $0.88$ \\
& NP2 & $0.78$ & $0.88$ & $1.35$ & $1.09$ & $1.05$ \\
\hline
\multirow{4}{*}{$16 \leq \ptchic < 18\GeV$} & P1 & $1.32$ & $0.89$ & $-$ & $0.89$ & $0.88$ \\
& P2 & $0.79$ & $0.88$ & $1.33$ & $1.08$ & $1.06$ \\
& NP1 & $1.32$ & $0.89$ & $-$ & $0.89$ & $0.88$ \\
& NP2 & $0.79$ & $0.88$ & $1.33$ & $1.08$ & $1.06$ \\
\hline
\multirow{4}{*}{$18 \leq \ptchic < 22\GeV$} & P1 & $1.30$ & $0.89$ & $-$ & $0.90$ & $0.89$ \\
& P2 & $0.79$ & $0.89$ & $1.31$ & $1.07$ & $1.06$ \\
& NP1 & $1.30$ & $0.89$ & $-$ & $0.90$ & $0.89$ \\
& NP2 & $0.79$ & $0.89$ & $1.31$ & $1.07$ & $1.06$ \\
\hline
\multirow{4}{*}{$22 \leq \ptchic < 30\GeV$} & P1 & $1.26$ & $0.90$ & $-$ & $0.90$ & $0.90$ \\
& P2 & $0.81$ & $0.90$ & $1.27$ & $1.06$ & $1.05$ \\
& NP1 & $1.26$ & $0.90$ & $-$ & $0.90$ & $0.90$ \\
& NP2 & $0.81$ & $0.90$ & $1.27$ & $1.06$ & $1.05$ \\
\hline
\end{tabular}
\end{center}

\caption{Scale factors that modify the central cross-section values, evaluated assuming isotropic decay angular distributions, to a given spin alignment scenario. The different spin alignment scenarios are defined in table\,\ref{Table:Pol}. The labels $\mathrm{(N)P1}$ and $\mathrm{(N)P2}$ correspond to (non-)prompt $\chicj1$ and (non-)prompt $\chicj2$ respectively.}
\label{Table:ScaleFactors}
\end{table}

\onecolumn
\clearpage 
\begin{flushleft}
{\Large The ATLAS Collaboration}

\bigskip

G.~Aad$^{\rm 48}$,
T.~Abajyan$^{\rm 21}$,
B.~Abbott$^{\rm 112}$,
J.~Abdallah$^{\rm 12}$,
S.~Abdel~Khalek$^{\rm 116}$,
O.~Abdinov$^{\rm 11}$,
R.~Aben$^{\rm 106}$,
B.~Abi$^{\rm 113}$,
M.~Abolins$^{\rm 89}$,
O.S.~AbouZeid$^{\rm 159}$
% ...
\bigskip
\\
% ...
$^{11}$ Institute of Physics, Azerbaijan Academy of Sciences, Baku, Azerbaijan\\
$^{12}$ Institut de F{\'\i}sica d'Altes Energies and Departament de F{\'\i}sica de la Universitat Aut{\`o}noma de Barcelona, Barcelona, Spain\\
% ...
$^{21}$ Physikalisches Institut, University of Bonn, Bonn, Germany\\
% ...
$^{48}$ Fakult{\"a}t f{\"u}r Mathematik und Physik, Albert-Ludwigs-Universit{\"a}t, Freiburg, Germany\\
% ...
$^{89}$ Department of Physics and Astronomy, Michigan State University, East Lansing MI, United States of America\\
% ...
$^{106}$ Nikhef National Institute for Subatomic Physics and University of Amsterdam, Amsterdam, Netherlands\\
% ...
$^{112}$ Homer L. Dodge Department of Physics and Astronomy, University of Oklahoma, Norman OK, United States of America\\
$^{113}$ Department of Physics, Oklahoma State University, Stillwater OK, United States of America\\
% ...
$^{116}$ LAL, Universit{\'e} Paris-Sud and CNRS/IN2P3, Orsay, France\\
% ...
$^{159}$ Department of Physics, University of Toronto, Toronto ON, Canada\\
% ...
\end{flushleft}

\end{document}